\documentclass[showpacs,preprintnumbers,
superscriptaddress,amsmath,floatfix]{revtex4}

\usepackage{pstcol}             
\usepackage{graphicx}
\definecolor{red}{rgb}{0.7,0,0}

\renewcommand\a{\alpha}

\newcommand\g{\gamma}
\renewcommand\d{\delta}
\newcommand\e{\epsilon}

\newcommand\m{\mu}

\newcommand\G{\Gamma}

\newcommand{\non}{\nonumber\\}

\newcommand{\be}{\begin{equation}}
\newcommand{\ee}{\end{equation}}
\newcommand{\bea}{\begin{eqnarray}}
\newcommand{\eea}{\end{eqnarray}}
\newcommand{\ba}[1]{\begin{array}{#1}}
\newcommand{\ea}{\end{array}}

\newcommand{\bm}[1]{\mbox{\boldmath${#1}$}}
 
\newcommand{\uk}{\hat{\mathbf{k}}}
\newcommand{\up}{\hat{\mathbf{p}}}
\newcommand{\hk}{\hat{k}}

\newcommand{\hp}{\hat{p}}
\newcommand{\vg}{\bm{\gamma}}
\newcommand{\vS}{\bm{\Sigma}}
\newcommand{\gperp}{\bm{\gamma}_{\perp}}

\newcommand{\Tr}{{\rm Tr}}

\begin{document}

\title{Neutrino emission and cooling rates of spin-one 
color superconductors} 

\author{Andreas Schmitt}
\email{aschmitt@lns.mit.edu}
\affiliation{Center for Theoretical Physics, Massachusetts Institute 
of Technology, Cambridge, MA 02139, USA}

\author{Igor A. Shovkovy}
\email{shovkovy@th.physik.uni-frankfurt.de}
\altaffiliation[on leave from ]{%
       Bogolyubov Institute for Theoretical Physics,
       03143, Kiev, Ukraine}
\affiliation{
Frankfurt Institute for Advanced Studies, J.W.\ Goethe-Universit\"at,
D-60054 Frankfurt am Main, Germany}

\author{Qun Wang}
\email{qunwang@ustc.edu.cn}
\affiliation{
Department of Modern Physics, University of Science and Technology of China,
Hefei, Anhui 230026, People's Republic of China}

\date{\today}

\begin{abstract}
Neutrino emissivities due to direct Urca processes of several spin-one color-superconducting 
phases of dense quark matter are calculated.  
In particular, the role of anisotropies and nodes of the gap functions is analyzed. 
Results for the specific heat as well as for the cooling rates of the color-spin-locked,
planar, polar, and {\it A} phases are presented and consequences for the physics of neutron stars
are briefly discussed. Furthermore, it is shown that the {\em A} phase exhibits a helicity order,
giving rise to a reflection asymmetry in the neutrino emissivity.

\end{abstract}

\maketitle

\section{Introduction}

Exotic states of matter are expected to exist in the central 
regions of compact stars. The baryon density in these systems 
is likely to exceed several times the nuclear saturation density,
$\rho_0\simeq 0.15~\mbox{fm}^{-3}$. The exact nature of such 
dense matter, however, is yet unknown. It was suggested already 
30 years ago \cite{perry} that it might be deconfined quark matter. 
Since the temperatures in neutron stars are sufficiently low, 
this matter is likely to be in a color-superconducting state. 
(For reviews on color superconductivity see Refs.~\cite{reviews}.) 
There is little doubt that dense baryonic matter at asymptotically 
high quark chemical potential $\mu$ is a color superconductor 
\cite{bailin}. In this case, the ground state is the color-flavor 
locked (CFL) phase \cite{alford} (for studies of the CFL phase in 
QCD at asymptotic density, see also Refs.~\cite{weak-cfl1,weak-cfl2}). 
At densities existing in stars, however, this phase may not 
necessarily be realized. The main reason for the potential breakdown of the 
CFL phase is a relatively large difference between the masses of 
the strange quark, $m_s$, and the masses of the up and down quarks, 
$m_u\simeq m_d$, which is negligible only at large densities. This 
difference, together with the requirements of $\beta$ equilibrium 
and electric and color charge neutrality, gives rise to a mismatch 
between the Fermi momenta of the quarks that form Cooper pairs
\cite{neutrality}. Therefore, the conventional BCS pairing \cite{bcs}, 
which is the underlying mechanism of the CFL state, becomes 
questionable and the search for the true ground state of quark 
matter in compact stars has to include different, unconventional 
superconducting states \cite{shovkovy,gCFL,cfl+mesons,alford2}. 

Determining the ground state of QCD at moderate densities from 
first principles, i.e., within the framework of QCD, is not possible 
at present. It is natural therefore to make use of astrophysical 
observations to test the presence of various suggested phases. 
The goal of this paper is to investigate color superconductors 
in which quarks of the same flavor form Cooper pairs 
and to study their effect on specific observables. 

Single-flavor Cooper pairing is the simplest possibility for neutral 
quark matter. Contrary to other unconventional pairing mechanisms, such 
as the gapless 2SC phase \cite{shovkovy}, the gapless CFL phase \cite{gCFL}, 
the CFL phase with additional meson condensates \cite{cfl+mesons}, 
or the crystalline phases \cite{alford2}, it is allowed for 
arbitrarily large mismatches between the Fermi momenta of different 
quark flavors. Single-flavor pairing in the color anti-triplet 
channel is possible only in the symmetric spin-one channel 
\cite{bailin,spin1,schaefer,Schmitt:2002sc,Schmitt:2003xq,andreas}. 
This is the consequence of the Pauli principle which requires 
the wave function of the Cooper pair to be antisymmetric under 
the exchange of the constituent quarks. This is in contrast 
to pairing of different flavors where the antisymmetric spin-zero 
channel is allowed. 

Furnishing triplet representations with respect to both color 
and spin groups, the order parameter in a spin-one color 
superconductor is given by a complex 3$\times$3 matrix. 
This is similar to superfluid $^3$He, where condensation 
occurs in spin and angular momentum triplets \cite{vollhardt}. 
In both systems, the matrix structure of the order parameter 
gives rise to several possible phases. In $^3$He the observed 
phases are the so-called {\it A}, {\it B}, and $A_1$ phases. 
In this paper, we consider the following four main spin-one 
color-superconducting phases of quark matter: the color-spin 
locked (CSL), planar, polar, and {\it A} phases, proposed in 
Refs.~\cite{schaefer,Schmitt:2002sc,Schmitt:2003xq,andreas,Buballa:2002wy}. 
The {\it A} and CSL phases are analogues of the {\it A} and 
{\it B} phases in $^3$He, respectively.  

In contrast to the nonrelativistic case of $^3$He, spin-one 
color-superconductors may involve pairing of quarks 
of the same as well as of opposite chiralities. In this 
paper, we focus on the so-called ``transverse'' phases, 
in which exclusively quarks of opposite chiralities pair. 
Theoretical studies at asymptotically large densities 
suggest that these phases are preferred 
\cite{schaefer,andreas}.

As in the case of $^3$He, the gap functions of most of the 
quark phases considered here are anisotropic in momentum 
space. Only the CSL phase is isotropic. The gap in the polar 
phase vanishes at the south and north poles of the Fermi 
sphere, whereas the gap in the planar phase is anisotropic 
but nonzero in any direction of the quasiparticle momentum. 
The {\it A} phase is special in the sense that it has two gapped 
quasiparticle modes with different angular structures, one 
of which has two point nodes. 

One should expect that the anisotropies and especially the nodes 
affect the physical properties of the corresponding quark phases. 
The low energy excitations around the nodes give important 
contributions to various thermodynamical and transport properties, 
e.g., the specific heat, the neutrino emissivity, the viscosity, 
the heat and electrical conductivity, etc. In application to compact 
stars, the specific heat and the neutrino emissivity determine the 
cooling behavior during the first $10^5$ -- $10^6$ years of the 
stellar evolution. In this paper, we compute these two quantities 
for the four mentioned spin-one color-superconducting phases and deduce the 
resulting cooling rates. In 
addition, we will discuss the reason why the distribution of neutrino 
emission from the {\it A} phase breaks reflection symmetry in position 
space.

The paper is organized as follows. The general formalism of 
calculating the time derivative of the neutrino distribution 
function is given in Sec.~\ref{formalism}. The main result
from Sec.~\ref{formalism} is then used in Sec.~\ref{results}
to compute the neutrino emissivity. The 
specific heat is calculated in Sec.~\ref{specificheat}. Both 
quantities are used in Sec.~\ref{cooling} to discuss the 
cooling behavior of the considered phases. In Sec.~\ref{kicks} 
we explain the asymmetry of the neutrino emission in the {\it A} 
phase. Finally, we comment briefly on the effect of nonzero 
quark masses in Sec.~\ref{quarkmass}.

Our convention for the metric tensor is 
$g^{\mu\nu}=\mbox{diag}(1,-1,-1,-1)$. 
Our units are $\hbar=c=k_B=1$. Four-vectors are denoted 
by capital letters, $K\equiv K^\mu=(k_0,{\bf k})$, while 
$k\equiv|{\bf k}|$ and $\uk\equiv{\bf k}/k$.
We work in the imaginary-time formalism, i.e., 
$T/V \sum_K \equiv T \sum_n \int d^3{\bf k}/(2\pi)^3$, 
where $n$ labels the Matsubara frequencies $\omega_n \equiv i k_0$. 
The Matsubara frequencies are $\omega_n\equiv 2n \pi T$ for bosons,
and $\omega_n\equiv(2n+1) \pi T$ for fermions.

\section{Time derivative of the neutrino distribution function}
\label{formalism}

In this section, we derive a general expression for the time derivative 
of the neutrino distribution function in spin-one color-superconducting 
phases. 

\subsection{General formalism}

Within the Kadanoff-Baym formalism \cite{KB}, one derives the following 
kinetic equation for the neutrino Green function (see for example 
Refs.~\cite{VoskSen,friman,kadanoff}):
\begin{equation}
i\partial_{X}^{\lambda}\mbox{Tr}[\gamma_{\lambda}G^{<}_{\nu}(X,P_\nu)]
=-\mbox{Tr}\left[G^{>}_{\nu}(X,P_\nu)\Sigma^{<}_{\nu}(X,P_\nu)
                  -\Sigma^{>}_{\nu}(X,P_\nu)G^{<}_{\nu}(X,P_\nu)\right] \,\, ,
\label{KB-eq}
\end{equation}
where the trace runs over Dirac space. (Here and in the following, 
the index $\nu$ always labels neutrino quantities and should not 
be confused with a Lorentz index.) The kinetic equation is obtained 
from the general Kadanoff-Baym equation after applying a 
gradient expansion, which is valid when the neutrino Green functions 
$G^{<,>}_\nu(X,P_\nu)$ and self-energies $\Sigma^{<,>}_\nu (X,P_\nu)$ are slowly 
varying functions of the space-time coordinate $X=(t,\mathbf{x})$. 
For our purposes, it is sufficient 
to consider spatially homogeneous systems which are close to equilibrium. 
In this case, the neutrino Green functions assume the following approximate 
form \cite{friman}:
\begin{subequations}
\label{Glesslarge}
\begin{eqnarray}
iG^{<}_{\nu}(t,P_\nu) & = & -(\gamma^\lambda P_{\nu,\lambda} +\mu_\nu\gamma_0 )
\frac{\pi}{p_\nu}\left\{ f_{\nu}(t,\mathbf{p}_\nu)\,\delta(p_{\nu 0}+\mu_{\nu}-p_\nu) 
-[1-f_{\overline{\nu}}(t,-\mathbf{p}_\nu)]\,\delta(p_{\nu 0}+\mu_\nu+p_\nu)
\right\} \, ,
\label{G-less1} \\ 
iG^{>}_{\nu}(t,P_\nu) & = & (\gamma^\lambda P_{\nu,\lambda} +\mu_\nu\gamma_0 )
\frac{\pi}{p_\nu}\left\{ [1-f_{\nu}(t,\mathbf{p}_\nu)] \,\delta(p_{\nu 0}+\mu_{\nu}-p_\nu)
 -f_{\overline{\nu}}(t,-\mathbf{p}_\nu) \,\delta(p_{\nu 0}+\mu_\nu +p_\nu)
\right\} \, ,
\label{G-larger1}
\end{eqnarray}
\end{subequations}
where $\mu_\nu$ is the neutrino chemical potential. The functions 
$f_{\nu,\overline{\nu}}(t,\mathbf{p}_\nu)$ are the neutrino 
and antineutrino distribution functions. We are interested in 
the Urca processes $u+e^{-}\to d+\nu$ (electron capture)
and $d\to u+e^{-}+\overline{\nu}$ ($\beta$-decay), which 
provide the dominant cooling mechanism for quark matter. 
These processes are important for neutron stars 
with core temperatures of the order of or smaller than several MeV. 
In this stage of the stellar evolution, the mean free path of neutrinos becomes larger than the 
stellar radius, wherefore in our results we shall set $\mu_\nu = 0$. 

\begin{figure}
\includegraphics[width=0.9\textwidth]{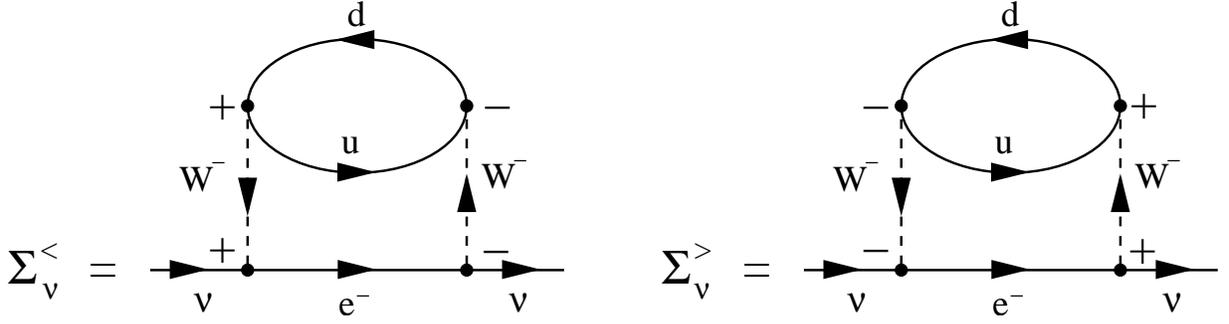}
\caption{\label{fig:urca}
Neutrino self-energies relevant for the 
neutrino Urca processes in the close-time-path formalism. 
The $+$ and $-$ signs assign the vertices to the 
upper and lower branch of the time contour, respectively. }
\end{figure}

The leading order contributions
to the neutrino self-energies which enter the kinetic equation 
(\ref{KB-eq}) are given by the diagrams in Fig.~\ref{fig:urca}. 
For the sake of simplicity, we do not take strange quarks into account. 
Their weak interactions are Cabibbo suppressed, and their number density is
not expected to be very large. (Admittedly, however, the bigger phase
space for the Urca processes involving massive strange quarks may 
partially compensate this suppression.) The diagrams in 
Fig.~\ref{fig:urca} translate into the following expressions,
\begin{subequations}
\label{eq3}
\begin{eqnarray}
\Sigma_{\nu}^{<}(t,P_\nu) & = &  
\frac{G_{F}^{2}}{2}\int\frac{d^{4}P_{e}}{(2\pi)^{4}}
\gamma^{\lambda}(1-\gamma_{5})
(\gamma^\kappa P_{e,\kappa} +\mu_e\gamma_0 )
\gamma^{\sigma}(1-\gamma_{5}) \non 
&&\times\;\Pi_{\lambda\sigma}^{>}(P_e-P_\nu)\frac{\pi}{p_{e}}f_{e}(t,\mathbf{p}_{e})
\delta(p_{e0}+\mu_{e}-p_{e}),\\
\Sigma_{\nu}^{>}(t,P_\nu) & = &  
-\frac{G_{F}^{2}}{2}\int\frac{d^{4}P_{e}}{(2\pi)^{4}}
\gamma^{\lambda}(1-\gamma_{5})
(\gamma^\kappa P_{e,\kappa} +\mu_e\gamma_0 )
\gamma^{\sigma}(1-\gamma_{5}) \non
&&\times\;\Pi_{\lambda\sigma}^{<}(P_e-P_\nu)\frac{\pi}{p_{e}}
[1-f_{e}(t,\mathbf{p}_{e})] \delta(p_{e0}+\mu_{e}-p_{e}),
\end{eqnarray}
\end{subequations}
where $G_F$ is the Fermi coupling constant and $\mu_e$ is the electron 
chemical potential. In the derivation we used the explicit form of 
the electron Green functions $G^{<,>}_{e}(t,P_e)$. They are given by expressions analogous
to Eqs.~(\ref{Glesslarge}). We have neglected the positron contribution, i.e., the analogues of the 
second terms in curly brackets on the right-hand sides of Eqs.~(\ref{Glesslarge}).
[Note that the processes involving 
positrons, $u\to d+e^{+}+\nu$ and $d+e^{+}\to u+\overline{\nu}$, 
are suppressed by a large factor $\exp(2\mu_e/T)$ because 
$\mu_e$ is positive and, in the regime under consideration, 
much larger than the temperature.]

The functions 
$\Pi_{\lambda\sigma}^{<,>}(P_e-P_\nu)$ are the self-energies 
of the $W$ bosons. The $W$ exchange is approximated
by its local form since the typical neutrino energies are much 
smaller than the $W$ mass. In the case of neutrino processes 
in compact stars, the corresponding neutrino 
energies do not exceed several dozens MeV. 

Next, we insert the Green functions (\ref{Glesslarge}) and self-energies (\ref{eq3}) into the kinetic
equation (\ref{KB-eq}). In order to obtain the expression 
for the neutrino (antineutrino)
distribution function, we integrate on both sides 
of the kinetic equation over $p_{\nu 0}$ from $-\mu_\nu$ to $\infty$ 
(from $-\infty$ to $-\mu_\nu$). The results are
\begin{subequations}
\label{df/dt-nu1}
\begin{eqnarray}
\frac{\partial}{\partial t}f_{\nu}(t,\mathbf{p}_\nu) & = & 
-i\frac{G_{F}^{2}}{16}\int\frac{d^{3}\mathbf{p}_{e}}{(2\pi)^{3}p_\nu p_e}
 L^{\lambda\sigma}({\bf p}_e,{\bf p}_\nu)
\,\left\{[1-f_{\nu}(t,\mathbf{p}_\nu)]\,f_{e}(t,\mathbf{p}_{e})
\,\Pi_{\lambda\sigma}^{>}(Q) \right.\non 
&& \left. \hspace{5cm} - \; f_{\nu}(t,\mathbf{p}_\nu)\,
[1-f_{e}(t,\mathbf{p}_{e})]\,\Pi_{\lambda\sigma}^{<}(Q)
\right\} \,\, , 
\label{df/dt-nu}\\
\frac{\partial}{\partial t}f_{\overline{\nu}}(t,\mathbf{p}_\nu) & = &
-i\frac{G_{F}^{2}}{16}\int\frac{d^{3}\mathbf{p}_{e}}{(2\pi)^{3}p_\nu p_e}
L^{\lambda\sigma}({\bf p}_e,{\bf p}_\nu)
\,\left\{[1-f_{\overline{\nu}}(t,\mathbf{p}_\nu)]
\,[1-f_{e}(t,\mathbf{p}_{e})]\,
\Pi_{\lambda\sigma}^{<}(Q^\prime) \right.\non
&& \left. \hspace{5cm} - \; f_{\overline{\nu}}(t,\mathbf{p}_\nu)
\,f_{e}(t,\mathbf{p}_{e})\,
\Pi_{\lambda\sigma}^{>}(Q^\prime)\right\}\,\, ,
\label{df/dt-nu-bar}
\end{eqnarray}
\end{subequations}
with the four-momenta $Q\equiv (p_e-p_\nu-\mu_e+\mu_\nu,{\bf p}_e-{\bf p}_\nu)$  
and $Q'\equiv (p_e+p_\nu-\mu_e+\mu_\nu,{\bf p}_e+{\bf p}_\nu)$, and the shorthand 
notation 
\begin{equation} \label{defL}
L^{\lambda\sigma}({\bf p}_e,{\bf p}_\nu)\equiv\mbox{Tr}\left[(\g_0p_e-\vg\cdot{\bf p}_e)\, 
\gamma^\sigma (1-\gamma^5)(\g_0p_\nu-\vg\cdot{\bf p}_\nu)\, \gamma^\lambda
(1-\gamma^5)\right].
\end{equation}

In the following, it is convenient to express the results in terms of the
retarded self-energy for $W$ bosons, $\Pi^{\lambda\sigma}_R$. 
Therefore, we shall use the following relations:
\begin{subequations}
\begin{eqnarray}
\Pi^{>}(Q) & = & -2i[1+n_{B}(q_{0})]\mbox{Im}\Pi_{R}(Q) \,\, , \\
\Pi^{<}(Q) & = & -2i\,n_{B}(q_{0})\mbox{Im}\Pi_{R}(Q) \,\, , 
\end{eqnarray}
\end{subequations}
where $n_{B}(\omega)\equiv 1/(e^{\omega/T}-1)$ is the 
Bose-Einstein distribution function. 

Here we consider the cooling of quark matter only 
in the absence of neutrino trapping. Then, the 
neutrino and antineutrino distribution functions on the right-hand side
of Eq.~(\ref{df/dt-nu1}) are vanishing, 
$f_{\nu,\overline{\nu}}(t,\mathbf{p}_\nu)=0$.
The electron distribution function can be approximated by its
equilibrium expression, 
\be
f_{e}(t,\mathbf{p}_e)\simeq n_F(p_e-\mu_e) \,\, ,
\ee
with $n_F(\omega)\equiv 1/(e^{\omega/T}+1)$. Then, one arrives at 
\begin{subequations}
\begin{eqnarray}
\frac{\partial}{\partial t}f_{\nu}(t,\mathbf{p}_\nu) & = & 
\frac{G_{F}^{2}}{8}\int\frac{d^{3}\mathbf{p}_{e}}{(2\pi)^{3}p_\nu p_e}
L_{\lambda\sigma}({\bf p}_e,{\bf p}_\nu)\,n_F(p_e-\mu_e) \,
n_{B}(p_\nu + \mu_e - p_e)
\,\mathrm{Im}\Pi_R^{\lambda\sigma}(Q), 
\label{df/dt-nu-1}\\
\frac{\partial}{\partial t}f_{\overline{\nu}}(t,\mathbf{p}_\nu) & = &
-\frac{G_{F}^{2}}{8}\int\frac{d^{3}\mathbf{p}_{e}}{(2\pi)^{3}p_\nu p_e}
L_{\lambda\sigma}({\bf p}_e,{\bf p}_\nu)\, 
n_F(\mu_e - p_e)\,n_{B}(p_\nu - \mu_e + p_e)
\,\mathrm{Im}\Pi_{R}^{\lambda\sigma}(Q').
\label{df/dt-nu-bar-1}
\end{eqnarray}
\end{subequations}
We shall see in Sec.~\ref{General-result} that the right-hand sides 
of Eqs.~(\ref{df/dt-nu-1}) and (\ref{df/dt-nu-bar-1}) are in fact identical.

\subsection{Quark propagators in spin-one color superconductors}

As we shall see below, cf.\ Eq.~(\ref{Pidef}), the expression for the imaginary 
part of the polarization tensor of the W-vector boson is given in terms of 
the quark propagator $S(K)$. In this paper, we consider the following
four spin-one superconducting phases: the CSL, planar, polar, and {\it A}
phases. Each of these phases is characterized by a $12\times 12$ gap matrix 
${\cal M}_{\bf k}$ in color and Dirac space,
\be
{\cal M}_{\bf k} = \sum_{i,j=1}^3J_i\Delta_{ij}\g_{\perp,j}(\uk) \,\, , 
\label{defM}
\ee  
where $(J_i)_{jk}=-i\e_{ijk}$ and 
$\gperp(\uk)\equiv \vg - \uk\,\bm{\g}\cdot\uk$ 
are basis vectors for the color antitriplet and spin triplet representations, 
respectively. We focus exclusively on the transverse phases, which have the 
highest pressure at asymptotic density and thus are expected to be preferred 
over others. For a more general form of the matrix ${\cal M}_{\bf k}$ see
Ref.~\cite{andreas}. The matrix $\Delta$ is a complex $3\times 3$ matrix 
and assumes a specific structure for each of the phases. 
In Table~\ref{tablephases}, we give $\Delta$ and the resulting matrices 
${\cal M}_{\bf k}$.

\begin{table}[t]
\begin{tabular}[t]{|c||c|c||c|c|c|} 
\hline
\;\; phase \;\; & $\Delta_{ij}$ & ${\cal M}_{\bf k}$ & 
\;\; $\lambda_{{\bf k},1}\;(n_1)$  \;\; & 
\;\; $\lambda_{{\bf k},2}\;(n_2)$  \;\; & 
\;\; $\lambda_{{\bf k},3}\;(n_3)$  \;\;  \\ 
\hline\hline
CSL & $\delta_{ij}$ & ${\bf J}\cdot\gperp(\uk)$ & 2\;(8) & 0\;(4) & --  \\
\hline  
planar & $\delta_{i1}\delta_{j1}+\delta_{i2}\delta_{j2}$ & $J_1\g_{\perp,1}(\uk) + J_2\g_{\perp,2}(\uk)$ & 
$1 + \cos^2\theta_{\bf k}\;(8)$ & 0\;(4) & --  \\ 
\hline
polar & $\delta_{i3}\delta_{j3}$ & $J_3\g_{\perp,3}(\uk)$ & $\sin^2\theta_{\bf k} \;(8)$& 0\;(4) & --  \\ 
\hline
{\it A} &\;\;  $\delta_{i3}(\delta_{j1} + i\,\delta_{j2})$\;\; & \;\; $J_3[\g_{\perp,1}(\uk) + i\,\g_{\perp,2}(\uk)]$\;\; 
& \;\;$(1 + |\cos\theta_{\bf k}|)^2\;(4)$ \;\;& 
\;\;$(1 - |\cos\theta_{\bf k}|)^2\;(4)$\;\; & 0 \;(4) \\ 
\hline
\end{tabular}
\caption{Matrices $\Delta$ and ${\cal M}_{\bf k}$ and eigenvalues $\lambda_{{\bf k},r}$ with 
corresponding degeneracies $n_r$
for four spin-one color superconductors. The angle between ${\bf k}$ and the 
$z$-axis is denoted by $\theta_{\bf k}$.}
\label{tablephases}
\end{table}

In spin-one color superconductors, the quark propagator is diagonal in flavor space, 
$S(K) = {\rm diag}[S_u(K),S_d(K)]$. The Nambu-Gorkov structure of the flavor-diagonal 
elements is given by 
\be \label{prop}
S_f(K) = \left(\begin{array}{cc} G_f^+(K) & \Xi_f^-(K) \\ 
\Xi_f^+(K) & G_f^-(K) \end{array}\right) \,\, , 
\qquad f=u,d \,\, ,
\ee
where \cite{Schmitt:2002sc,Schmitt:2003xq,andreas}
\be \label{prop-diag}
G^\pm_f(K) = \left[G_{0,f}^\mp(K)\right]^{-1}\,\sum_{e,r}
\frac{{\cal P}_{{\bf k},r}^\pm\, \Lambda_{\bf k}^{\mp e}}
{k_0^2 - (\e_{{\bf k},r,f}^e)^2} \,\, .
\ee
For the sake of simplicity, we have not included the quark self-energy correction. For a more
general expression of the propagator, including this correction, see 
Refs.~\cite{Brown:2000eh,Wang:2001aq}. The inverse 
free propagator for quarks and charge-conjugate quarks in the ultrarelativistic 
limit is
\be \label{freeprop}
\left[G_{0,f}^\pm(K)\right]^{-1} = \g^\m K_\m\pm\m_f\g_0 = \g_0\sum_e[k_0\pm 
(\mu_f -ek)]\,\Lambda_{\bf k}^{\pm e}\,\, .
\ee
The Dirac matrices $\Lambda_{\bf k}^e\equiv (1+e\g_0\vg\cdot\uk)/2$, where $e=\pm$, 
are projectors onto positive and negative energy states. The matrices 
${\cal P}_{{\bf k},r}^{-}$ and ${\cal P}_{{\bf k},r}^{+}$ in Eq.\ (\ref{prop-diag}) are projectors onto 
the eigenspaces of the matrices ${\cal M}_{\bf k}{\cal M}^{\dagger}_{\bf k}$
and $\gamma^0{\cal M}^{\dagger}_{\bf k}{\cal M}_{\bf k}\gamma^0$, respectively.
Both matrices have the same set of eigenvalues $\lambda_{{\bf k},r}$,
\bea
{\cal M}_{\bf k}{\cal M}^{\dagger}_{\bf k} &\equiv &
\sum_{r} \lambda_{{\bf k},r}{\cal P}_{{\bf k},r}^{-} \,\, , \\
\gamma^0{\cal M}^{\dagger}_{\bf k}{\cal M}_{\bf k}\gamma^0 &\equiv &
\sum_{r} \lambda_{{\bf k},r}{\cal P}_{{\bf k},r}^{+} \,\, .
\eea
As we shall see,
only the projection operators ${\cal P}_{{\bf k},r}^{+}$ (and not 
${\cal P}_{{\bf k},r}^{-}$) are needed in the 
calculation of the neutrino emission. They are given explicitly in Appendix 
\ref{colordirac}. The eigenvalues $\lambda_{{\bf k},r}$ 
appear in the quasiparticle dispersion relations,
\be \label{excite}
\e_{{\bf k},r,f}^e = \sqrt{(ek-\mu_f)^2 + \lambda_{{\bf k},r}|\phi_f|^2} \,\, .
\ee
Here $\phi_f$ are the gap parameters, which are different for $u$ and $d$ quarks 
in general. The eigenvalues for the four considered phases as well as their 
degeneracies $n_{r}$ are listed in Table \ref{tablephases}. Note that all phases contain an ungapped
mode, $\lambda_{{\bf k},2(3)} = 0$.  

The off-diagonal elements on the right-hand side of Eq.~(\ref{prop}) 
are the so-called anomalous propagators. They are given by
\begin{subequations} \label{anomalous}
\bea \label{S212SC}
\Xi^+_f(K)&=&-\sum_{e,r} \gamma_0 \, {\cal M}_{\bf k}\, 
\gamma_0\, {\cal P}_{{\bf k},r}^+ \Lambda_{\bf k}^{-e}
\, \frac{\phi_f}{k_0^2-
(\e_{{\bf k},r,f}^e)^2} \,\, , \\
\Xi^-_f(K)&=&-\sum_{e,r} {\cal M}_{\bf k}^\dag\, 
 {\cal P}_{{\bf k},r}^- \Lambda_{\bf k}^e
\, \frac{\phi^*_f}{k_0^2-
(\e_{{\bf k},r,f}^e)^2} \,\, .
\eea
\end{subequations}
The propagators in Eq.~(\ref{prop-diag}) have particle and hole type 
poles at $k_0=\pm \e_{{\bf k},r,f}^+$, as well as the corresponding 
antiparticle poles at $k_0=\pm \e_{{\bf k},r,f}^-$. In the calculation 
of the imaginary part of the retarded self-energy $\Pi^{\lambda\sigma}_R(Q)$, 
the antiparticle contributions are suppressed by inverse powers of 
the quark chemical potential. [Note, however, that taking 
antiparticles into account is important in the calculation 
of $\mbox{Re}\Pi^{\lambda\sigma}_R(Q)$.] In our calculation, 
therefore, we omit the terms with $e=-$ in the quark propagator 
and arrive at the following approximate form
\be \label{fullprop}
G^\pm_f(K) \simeq \g_0\,\Lambda_{\bf k}^{\mp}\,
\sum_r{\cal P}_{{\bf k},r}^\pm\, \frac{k_0\mp(\mu_f-k)}
{k_0^2 - \e_{{\bf k},r,f}^2} \,\, ,
\ee
where we denoted $\e_{{\bf k},r,f}\equiv\e_{{\bf k},r,f}^+$.

\subsection{Imaginary part of the W-boson polarization tensor}

In this subsection, we evaluate $\mathrm{Im}\Pi_R^{\lambda\sigma}$. 
As can be seen from Fig.~\ref{fig:urca}, the $W$-boson polarization
tensor can be written as
\bea 
\Pi^{\lambda\sigma}(Q) &\equiv & \frac{T}{V}
\sum_K \Tr[\G_-^\lambda\,S(K)\,\G_+^\sigma\,S(K+Q)]  \,\, ,
\label{Pidef}
\eea
where the trace runs over Nambu-Gorkov, color, flavor, and Dirac space. 
The 2$\times$2 Nambu-Gorkov structure of the vertices reads 
\be
\Gamma^\lambda_{\pm} = \left(\begin{array}{@{\extracolsep{2mm}}cc}
\gamma^\lambda(1-\gamma^5)\,\tau_{\pm} & 0 \\
0 & -\gamma^\lambda (1+\gamma^5)\,\tau_{\mp}
\end{array}\right) \, ,
\label{vert}
\ee
where $\tau_{\pm}\equiv(\tau_1\pm i\tau_2)/2$ are matrices in flavor 
space, constructed from the Pauli matrices $\tau_1$, $\tau_2$. 
After performing the traces over Nambu-Gorkov and flavor space, 
\be 
\label{Pi1} 
\Pi^{\lambda\sigma}(Q) = 
\frac{T}{V}\sum_K 
\left\{\Tr[\gamma^\lambda(1-\gamma^5)\,G_u^+(K)\,
\gamma^\sigma (1-\gamma^5)\, G^+_d(P)] +\Tr[\gamma^\lambda(1+\gamma^5)\,G_d^-(K)\,
\gamma^\sigma (1+\gamma^5)\, G^-_u(P)] \right\}  \,\, ,
\ee
where $P\equiv K+Q$, and the traces run over color and Dirac space. 
As we see, the anomalous propagators $\Xi^\pm_f$ do not contribute to 
the self-energy. This is in contrast to spin-zero color superconductors, such as 
the (gapless) 2SC and CFL phases, in which the anomalous propagators are 
off-diagonal in flavor space. In fact, this difference is related to the conservation 
of electric charge in the diagrams in Fig.~\ref{fig:urca}. The anomalous 
propagators contain the Cooper pair condensate, which, in the case of a spin-one color
superconductor, is of the form $\langle u u \rangle$ or $\langle d d  \rangle$. Therefore, 
the appearance of the anomalous propagators in the diagrams of Fig.~\ref{fig:urca} 
is forbidden by electric charge conservation.
In a spin-zero color superconductor, however, the condensate is of the form $\langle u d \rangle$.
Hence, electric charge can be extracted from or deposited into the condensate of Cooper 
pairs in the ground state and diagrams containing the anomalous propagators contribute to the 
$W$-boson polarization tensor \cite{jaikumar}.

After inserting Eq.~(\ref{fullprop}) into Eq.~(\ref{Pi1}), one arrives
at the following expression for the imaginary part,
\be 
\label{beforematsu}
{\rm Im}\Pi^{\lambda\sigma}_R(Q) = 
\frac{T}{V}{\rm Im}\,\sum_K
\sum_{r,s} \left\{ \frac{[k_0-(\mu_u - k)]\,[p_0-(\mu_d-p)]}{(k_0^2-
\e_{{\bf k},r,u}^2)\,(p_0^2-
\e_{{\bf p},s,d}^2)}\,{\cal T}_{rs,+}^{\lambda\sigma}(\uk,\up) 
+ \frac{[k_0+(\mu_d - k)]\,[p_0+(\mu_u-p)]}{(k_0^2-
\e_{{\bf k},r,d}^2)\,(p_0^2-
\e_{{\bf p},s,u}^2)}\,{\cal T}_{rs,-}^{\lambda\sigma}(\uk,\up) \right\}  ,
\ee
where we defined the following traces in color and Dirac space,
\be 
\label{defT}
{\cal T}_{rs,\pm}^{\lambda\sigma}(\uk,\up)\equiv \Tr\left[
\gamma^\lambda(1\mp\gamma^5)\,\g^0\,{\cal P}_{{\bf k},r}^\pm
\,\Lambda_{\bf k}^{\mp}\,\gamma^\sigma (1\mp\gamma^5)\,\g^0\,
{\cal P}_{{\bf p},s}^\pm\,\Lambda_{\bf p}^{\mp}\right]
\,\, .
\ee
In order to perform the 
Matsubara sum, we use Eq.\ (\ref{eqA4}) in Appendix~\ref{matsubara}. Then, extracting the imaginary part yields
\bea \label{Pi2}
{\rm Im}\Pi^{\lambda\sigma}_R(Q) 
&=& 
-\pi \sum_{r,s}\sum_{e_1,e_2=\pm}\int
\frac{d^3{\bf k}}{(2\pi)^3}
\left[{\cal T}_{rs,+}^{\lambda\sigma}(\uk,\up)\,
B_{{\bf k},r,u}^{e_1}\,B_{{\bf p},s,d}^{e_2}   \,
\frac{n_F(-e_1\e_{{\bf k},r,u})\,n_F(e_2\e_{{\bf p},s,d})}
{n_B(-e_1\e_{{\bf k},r,u}+e_2\e_{{\bf p},s,d})} \, 
\d(q_0-e_1\e_{{\bf k},r,u}+e_2\e_{{\bf p},s,d}) \right.\non
&& \left. +\;
{\cal T}_{rs,-}^{\lambda\sigma}(\uk,\up)\, 
B_{{\bf k},r,d}^{e_1}\,B_{{\bf p},s,u}^{e_2}  \,\frac{n_F(e_1\e_{{\bf k},r,d})\,
n_F(-e_2\e_{{\bf p},s,u})}{n_B(e_1\e_{{\bf k},r,d}-e_2\e_{{\bf p},s,u})}\,
\d(q_0+e_1\e_{{\bf k},r,d}-e_2\e_{{\bf p},s,u}) \right] \,\, .
\eea
Here, we defined the Bogoliubov coefficients
\be
\label{bglbv}
B_{{\bf k},r,f}^{e}\equiv \frac{\e_{{\bf k},r,f}+ e\,(\mu_f-k)}{2\e_{{\bf k},r,f}}
\,\, , \qquad (f=u,d \,\,, \quad e=\pm) \,\, .
\ee  
The two terms on the right-hand side of Eq.~(\ref{Pi2}) yield the same 
contribution. The physical reason for this is that the second term is just the 
charge-conjugate counterpart of the first term. The formal proof 
goes as follows. In the second term, one changes the summation indices 
$e_1\leftrightarrow e_2$ and $r\leftrightarrow s$. Then, one introduces 
the new integration variable ${\bf k} \to -{\bf k}-{\bf q}$
and uses 
\be
{\cal T}_{sr,-}^{\lambda\sigma}(-\up,-\uk) ={\cal T}_{rs,+}^{\lambda\sigma}(\uk,\up) 
\,\, ,
\ee
which holds for all phases considered in this paper.
After taking into account that $\lambda_{{\bf k},r}=\lambda_{-{\bf k},r}$, 
one obtains the first term in Eq.~(\ref{Pi2}). Consequently, in the following 
we keep only the first term and double the result, 
\bea 
{\rm Im}\Pi^{\lambda\sigma}_R(Q) &=& -2\pi\sum_{r,s}\sum_{e_1,e_2=\pm}\int
\frac{d^3{\bf k}}{(2\pi)^3} {\cal T}_{rs,+}^{\lambda\sigma}(\uk,\up)\,
B_{{\bf k},r,u}^{e_1}\,B_{{\bf p},s,d}^{e_2} \non
&&\times\,\frac{n_F(-e_1\e_{{\bf k},r,u})\,n_F(e_2\e_{{\bf p},s,d})}
{n_B(-e_1\e_{{\bf k},r,u}+e_2\e_{{\bf p},s,d})} \, 
\d(q_0-e_1\e_{{\bf k},r,u}+e_2\e_{{\bf p},s,d}) \,\, .
\label{pi-nu}
\eea 
The same result may also be written in another form which is more convenient 
for the use in Eq.~(\ref{df/dt-nu-bar-1}), i.e.,
\bea
{\rm Im}\Pi^{\lambda\sigma}_R(Q') &=& 2\pi\sum_{r,s}\sum_{e_1,e_2=\pm}\int
\frac{d^3{\bf k}}{(2\pi)^3} {\cal T}_{rs,+}^{\lambda\sigma}(\uk,\up^\prime)\,
B_{{\bf k},r,u}^{e_1}\,B_{{\bf p}^\prime,s,d}^{e_2}  \non 
&&\times \,\frac{n_F(e_1\e_{{\bf k},r,u})\,n_F(-e_2\e_{{\bf p}^\prime,s,d})}
{n_B(e_1\e_{{\bf k},r,u}-e_2\e_{{\bf p}^\prime,s,d})} \, 
\d(q'_0-e_1\e_{{\bf k},r,u}+e_2\e_{{\bf p}^\prime,s,d}) \,\,.
\label{pi-anti-nu}
\eea
where we have used Eq.~(\ref{A2}) and defined $P^\prime\equiv K + Q^\prime$.

\subsection{General result for the time derivative of 
the neutrino distribution function}
\label{General-result}

We now insert the results for the polarization tensors (\ref{pi-nu}) and 
(\ref{pi-anti-nu}) into Eqs.~(\ref{df/dt-nu-1}) and (\ref{df/dt-nu-bar-1}), respectively. 
In order to calculate $L_{\lambda\sigma}({\bf p}_e,{\bf p}_\nu)\,{\rm Im}\Pi_R^{\lambda\sigma}(Q)$,
we have to compute the contraction of the tensor $L_{\lambda\sigma}({\bf p}_e,{\bf p}_\nu)$ 
with the color-Dirac trace ${\cal T}_{rs,+}^{\lambda\sigma}(\uk,\up)$. 
In all cases considered in this paper, we can write the result as 
\be \label{contract}
L_{\lambda\sigma}({\bf p}_e,{\bf p}_\nu)\,{\cal T}_{rs,+}^{\lambda\sigma}(\uk,\up)
= 64\,(p_e-{\bf p}_e\cdot\uk)\,(p_\nu - {\bf p}_\nu\cdot\up)\,\omega_{rs}(\uk,\up) \,\, ,
\ee
where the functions $\omega_{rs}(\uk,\up)$ depend on the specific phase. They are calculated in 
Appendix \ref{colordirac}. 

Then, Eqs.~(\ref{df/dt-nu-1}) and (\ref{df/dt-nu-bar-1}) become
\begin{subequations} \label{beforeFermi}
\bea 
\frac{\partial }{\partial t}f_\nu(t,{\bf p}_\nu) &=& -16\,\pi\,G_F^2\,
\int\frac{d^3{\bf p}_e}{(2\pi)^3p_\nu p_e}
\int\frac{d^3{\bf k}}{(2\pi)^3} \,n_F(p_e-\mu_e)\,
\sum_{e_1,e_2=\pm}\sum_{rs} (p_e - {\bf p}_e\cdot\uk)\,(p_\nu - {\bf p}_\nu\cdot\up)\non
&&\times \;\omega_{rs}(\uk,\up)
\, B_{{\bf k},r,u}^{e_1}\,B_{{\bf p},s,d}^{e_2}\,
n_F(-e_1\e_{{\bf k},r,u})\,n_F(e_2\e_{{\bf p},s,d})\,\d(p_e-\mu_e-p_\nu-e_1\e_{{\bf k},r,u}
+e_2\e_{{\bf p},s,d}) \,\, ,
\label{nu} \\
\frac{\partial }{\partial t}f_{\overline{\nu}}(t,{\bf p}_\nu) &=& -16\,\pi\,G_F^2\,
\int\frac{d^3{\bf p}_e}{(2\pi)^3p_\nu p_e}
\int\frac{d^3{\bf k}}{(2\pi)^3} \,n_F(\mu_e-p_e)\,
\sum_{e_1,e_2=\pm}\sum_{rs} (p_e - {\bf p}_e\cdot\uk)\,(p_\nu - {\bf p}_\nu\cdot\up^\prime)\non
&&\times \;\omega_{rs}(\uk,\up^\prime)
\, B_{{\bf k},r,u}^{e_1}\,B_{{\bf p}^\prime,s,d}^{e_2}\,
n_F(e_1\e_{{\bf k},r,u})\,n_F(-e_2\e_{{\bf p}^\prime,s,d})\,\d(p_e-\mu_e+p_\nu-e_1\e_{{\bf k},r,u}
+e_2\e_{{\bf p}^\prime,s,d}) \,\, .
\label{nu-bar}
\eea
\end{subequations}
At this point it is appropriate to recall the known result that 
the neutrino emission from low-temperature weakly interacting 
matter of massless quarks is strongly suppressed by the 
kinematics of the Urca processes \cite{iwamoto}. In particular, 
energy and momentum conservation requires nearly 
collinear momenta of the participating electron, up quark 
and down quark (after taking into account that $p_{\nu} \sim
T \ll \mu_e,\mu_u,\mu_d$). Strong interaction between 
quarks changes the situation dramatically \cite{iwamoto}.
In this case, applying Landau's theory of Fermi liquids, 
the quark Fermi velocity is reduced, $v_{F}\simeq 1-\kappa$, 
where $\kappa \equiv 2\a_s/(3\pi)$ with the strong coupling 
constant $\alpha_s$. (Here we ignore non-Fermi liquid 
corrections. For their possible effects on the Urca processes
in ungapped nuclear or quark matter see Refs.~\cite{voskresensky} 
and \cite{schwenzer}, respectively.) A rigorous treatment of the 
Fermi liquid correction would require a nonzero quark self-energy, 
which we have omitted, see Eq.~(\ref{prop-diag}) and remark 
below that equation. However, we may introduce the modified 
dispersion relations by hand in order to reproduce the result 
for normal quark matter as a limit case of our expressions. 
To this end, we identify the first three factors (multiplied 
by $k\, p$) on the right hand side of Eq.~(\ref{contract}) 
with the squared scattering amplitude $|M|^2$, derived by 
Iwamoto \cite{iwamoto},
\begin{equation}
64 G_F^2 (p_{e}k-\mathbf{p}_{e}\cdot\mathbf{k})
(p_{\nu}p-\mathbf{p}_{\nu}\cdot\mathbf{p}) \to 
|M|^2 \equiv 64 G_F^2 (P_{e}\cdot K)(P_{\nu}\cdot P)\,.
\label{eq:vuc}
\end{equation}
After this replacement, we take the Fermi liquid corrections 
into account by simply following the same steps as in 
Ref.~\cite{iwamoto}. As in Ref.~\cite{iwamoto}, we work to 
lowest order in $\alpha_s$. Therefore, the results of this 
paper are valid, strictly speaking, only at densities much 
higher than in the interior of neutron stars. However, the
role of the strong interaction in the neutrino emission is 
merely to open a phase space for the weak processes. In view 
of this, the limitation due to uncontrollable strong interaction 
may not be so essential for understanding the qualitative 
features of the neutrino processes in quark matter at 
realistic densities.

We approximate the $\delta$-functions in 
Eqs.~(\ref{beforeFermi}) as follows. By making use of the 
definitions ${\bf p}\equiv\mathbf{k}+\mathbf{p}_e-\mathbf{p}_\nu$ 
and ${\bf p}^\prime\equiv\mathbf{k}+\mathbf{p}_e+\mathbf{p}_\nu$, 
it is easy to show that the arguments of the $\delta$-functions
vanish only if the angle between the momenta of up and down quarks
$\theta_{ud}$ is equal to a fixed value $\theta_0$, up to 
corrections suppressed by powers of $p_\nu/\mu_e$. The value
of the angle $\theta_0$ is given by 
$\cos\theta_{0} \equiv 1 - \kappa\mu_e^2/(\mu_u\mu_d)$. 
We note that $\theta_0$ is independent of 
the neutrino energy $p_\nu$. Then, both $\delta$-functions 
in Eqs.~(\ref{nu}) and (\ref{nu-bar}) are replaced by 
$\mu _e/(\mu_u\mu_d)\,\delta(\cos\theta_{ud}-\cos\theta_0)$. 
Making use of this fact, we rewrite the equations in the 
following form,
\begin{subequations} \label{fapprox12}
\bea 
\label{fapprox} 
\frac{\partial }{\partial t}f_\nu(t,{\bf p}_\nu) 
&\simeq& -\frac{64}{3}\,\alpha_s G_F^2\,
\mu_e\mu_u\mu_d\int\frac{dp\,d\Omega_{\bf p}}{(2\pi)^3}
\int\frac{dk\,d\Omega_{\bf k}}{(2\pi)^3} \,
(1-\cos\theta_{\nu d})\, \d(\cos\theta_{ud}-\cos\theta_0) \non
&& \times \sum_{e_1,e_2=\pm}\sum_{rs} \omega_{rs}(\uk,\up)
\, B_{{\bf k},r,u}^{e_1}\,B_{{\bf p},s,d}^{e_2}\,
n_F(-e_1\e_{{\bf k},r,u})\,n_F(e_2\e_{{\bf p},s,d})\,
n_F(p_\nu+e_1\e_{{\bf k},r,u}-e_2\e_{{\bf p},s,d}) \,\, ,
 \\
\frac{\partial }{\partial t}f_{\overline{\nu}}(t,{\bf p}_\nu) 
&\simeq& -\frac{64}{3}\,\alpha_s G_F^2\,
\mu_e\mu_u\mu_d\int\frac{dp\,d\Omega_{\bf p}}{(2\pi)^3}
\int\frac{dk\,d\Omega_{\bf k}}{(2\pi)^3} \,
(1-\cos\theta_{\nu d})\,
\d(\cos\theta_{ud}-\cos\theta_0) \non
&& \times \sum_{e_1,e_2=\pm}\sum_{rs} \omega_{rs}(\uk,\up)
\, B_{{\bf k},r,u}^{e_1}\,B_{{\bf p},s,d}^{e_2}\,
n_F(e_1\e_{{\bf k},r,u})\,n_F(-e_2\e_{{\bf p},s,d})\,
n_F(p_\nu-e_1\e_{{\bf k},r,u}+e_2\e_{{\bf p},s,d}) \,\, ,
\label{fapprox1}
\eea
\end{subequations}
where $\theta_{\nu d}$ is the angle between the three-momenta 
of the neutrino and the $d$ quark. Here we have changed the 
integration variable ${\bf p}_e$ to 
${\bf p}\equiv\mathbf{k}+\mathbf{p}_e-\mathbf{p}_\nu$ and 
${\bf p}^\prime\equiv\mathbf{k}+\mathbf{p}_e+\mathbf{p}_\nu$, 
respectively, and afterwards dropped the prime of
${\bf p}^\prime$ in Eq.~(\ref{fapprox1}).

Instead of dimensionful momenta, it is convenient to introduce the 
following three dimensionless variables,
\be
x \equiv \frac{p-\mu_d}{T} \,\, , \qquad
y \equiv \frac{k-\mu_u}{T} \,\, , \qquad 
v \equiv \frac{p_\nu}{T} \,\, . 
\ee
In terms of the variables $x$ and $y$, the range of integration 
runs from $-\mu_{u,d}/T$ to $\infty$. Since the main contribution 
in the integral comes from $x,y\ll \mu_{u,d}/T$, the results do 
not change if the lower boundary is extended to $-\infty$. Then, 
we can drop the $x$- and $y$-odd contributions from the Bogoliubov 
coefficients,
\begin{subequations}
\bea
B_{{\bf p},s,d}^{e_2} &=& \frac{1}{2}- 
\frac{e_2\, x}{2 \sqrt{x^2+\lambda_{{\bf p},s}\varphi_d^2}}
\,\, , \\  
B_{{\bf k},r,u}^{e_1} &=& \frac{1}{2}- 
\frac{e_1 \,y}{2 \sqrt{y^2+\lambda_{{\bf k},r}\varphi_u^2}}\, ,
\eea
\end{subequations}
in the integrand, and keep only the even contributions, i.e., 
a constant term $1/2$. After this is taken into account, we 
restrict the integration over $x$ and $y$ from $0$ to $\infty$.
Accounting for the integration from $-\infty$ to $0$ produces 
extra factors of 2 for each integration which compensate the 
factors of $1/2$ from the two Bogoliubov coefficients. At this
point, we notice that the difference between Eqs.~(\ref{fapprox}) 
and (\ref{fapprox1}) lies only in the signs in front of $e_1$ 
and $e_2$. Because of the summation over $e_1$ and $e_2$, the 
right-hand sides of both Eqs.~(\ref{fapprox}) and (\ref{fapprox1}) 
are identical.

Hence, the result can be written in the following approximate form,
\bea \label{varchange}
\frac{\partial}{\partial t}  f_\nu(t,{\bf p}_{\nu}) &=& 
\frac{\partial}{\partial t}  f_{\overline{\nu}}(t,{\bf p}_{\nu}) \non
&\simeq& -\frac{64}{3}\,\alpha_s G_F^2\,
\mu_e\mu_u\mu_d\,T^2 \sum_{rs} \int\frac{d\Omega_{\bf p}}{(2\pi)^3}
\int\frac{d\Omega_{\bf k}}{(2\pi)^3} \,
(1-\cos\theta_{\nu d})\,
\d(\cos\theta_{ud}-\cos\theta_0)  \, F_{\varphi_u\varphi_d}^{rs}(\uk,\up,v) \,\, ,
\eea
where 
\bea \label{Frs}
F_{\varphi_u\varphi_d}^{rs}(\uk,\up,v)&\equiv& 
\omega_{rs}(\uk,\up) \sum_{e_1,e_2=\pm}\int_0^\infty \int_0^\infty 
\,dx\,dy
\left(e^{-e_1\sqrt{y^2+\lambda_{{\bf k},r}\varphi_u^2}}+ 1\right)^{-1}
\left(e^{e_2\sqrt{x^2+\lambda_{{\bf p},s}\varphi_d^2}} + 1\right)^{-1} \non 
&&\hspace{5cm}\times\; \left(e^{v+e_1\sqrt{y^2+\lambda_{{\bf k},r}\varphi_u^2} - 
e_2\sqrt{x^2+\lambda_{{\bf p},s}\varphi_d^2}} + 1\right)^{-1}
\eea
with $\varphi_f\equiv \phi_f/T$. 

Three out of four angular integrations in Eq.~(\ref{varchange}) can be calculated
approximately in an analytical form.
Details of this calculation are deferred to Appendix \ref{angint}. We obtain the following 
main result of this section,
\be \label{resultgen}
\frac{\partial }{\partial t} f_\nu(t,{\bf p}_\nu)= 
\frac{\partial}{\partial t}  f_{\overline{\nu}}(t,{\bf p}_{\nu})\simeq -\frac{4\alpha_s G_F^2}{3\pi^4}\,
\mu_e\mu_u\mu_d\,T^2 \sum_r \int_{-1}^1 d\xi\,(1-\xi\cos\theta_\nu)\,
F_{\varphi_u\varphi_d}^{rr}(\xi,v) \,\, .
\ee
It holds for isotropic phases as well as for phases in which the order parameter picks a special
direction in momentum space, identified with the $z$ direction. We denote the angle between the 
neutrino momentum and the $z$-axis by $\theta_\nu$. The function $F_{\varphi_u\varphi_d}^{rr}(\xi,v)$ is 
obtained from the function $F_{\varphi_u\varphi_d}^{rs}(\uk,\up,v)$ in the collinear limit ($\uk=\up$). Its
explicit form reads 
\bea\label{Frr}
F_{\varphi_u\varphi_d}^{rr}(\xi,v) &=&\omega_{rr}(\xi) 
\sum_{e_1,e_2=\pm}\int_0^\infty \int_0^\infty 
\,dx\,dy
\left(e^{-e_1\sqrt{y^2+\lambda_{\xi,r}\varphi_u^2}}+ 1\right)^{-1}
\left(e^{e_2\sqrt{x^2+\lambda_{\xi,r}\varphi_d^2}} + 1\right)^{-1} \non 
&&\hspace{5cm}\times\; \left(e^{v+e_1\sqrt{y^2+\lambda_{\xi,r}\varphi_u^2} - 
e_2\sqrt{x^2+\lambda_{\xi,r}\varphi_d^2}} + 1\right)^{-1} \, .
\eea
Here $\xi\equiv \cos\theta_u=\cos\theta_d$ with $\theta_u$ being the angle between the 
$z$-axis and the $u$ quark momentum ${\bf k}$, and $\theta_d$  being the angle between 
the $z$-axis and the $d$ quark momentum ${\bf p}$. The functions $\omega_{rr}(\xi)$ and 
$\lambda_{\xi,r}$ are given in Table~\ref{tableomega}. One should note that all 
components of $\omega_{rs}(\uk,\up)$ with $r\neq s$ disappear in the collinear limit. 
This can be seen directly from their explicit expressions given in 
Appendix~\ref{colordirac}. Because of this property, every excitation branch $r$ 
yields a separate contribution to the result (\ref{resultgen}).

\begin{table}[t]
\begin{tabular}[t]{|c||c|c||c|c||c|c|} 
\hline
\;\; phase \;\;  & $\omega_{11}(\xi)$ & 
$\lambda_{\xi,1}$ &
$\omega_{22}(\xi)$ &
$\lambda_{\xi,2}$& 
\;\;$\omega_{33}(\xi)$ \;\; &
$\lambda_{\xi,3}$
\\ \hline\hline
CSL & 2 & 2 & 1 & 0 & -- & -- \\ \hline 
planar & 2 & $1+\xi^2$  &1 & $0$  & -- & --   \\ \hline
polar & 2 &  $1-\xi^2$  & 1 & $0$  & -- & -- \\ \hline
{\it A} & \;\;$1+{\rm sgn}(\xi)$\;\; & \;\;$(1+|\xi|)^2$ \;\; &
\;\;$1-{\rm sgn}(\xi)$ \;\;& \;\;$(1-|\xi|)^2$  \;\;&
1  & $0$ \\ \hline   
\end{tabular}
\caption{Functions $\omega_{rr}(\xi)$ and $\lambda_{\xi,r}$ 
for four spin-one color superconductors.}
\label{tableomega}
\end{table}

\section{Neutrino emissivity}
\label{results}

\begin{figure} [t]
\begin{center}
\hbox{\includegraphics[width=0.49\textwidth]{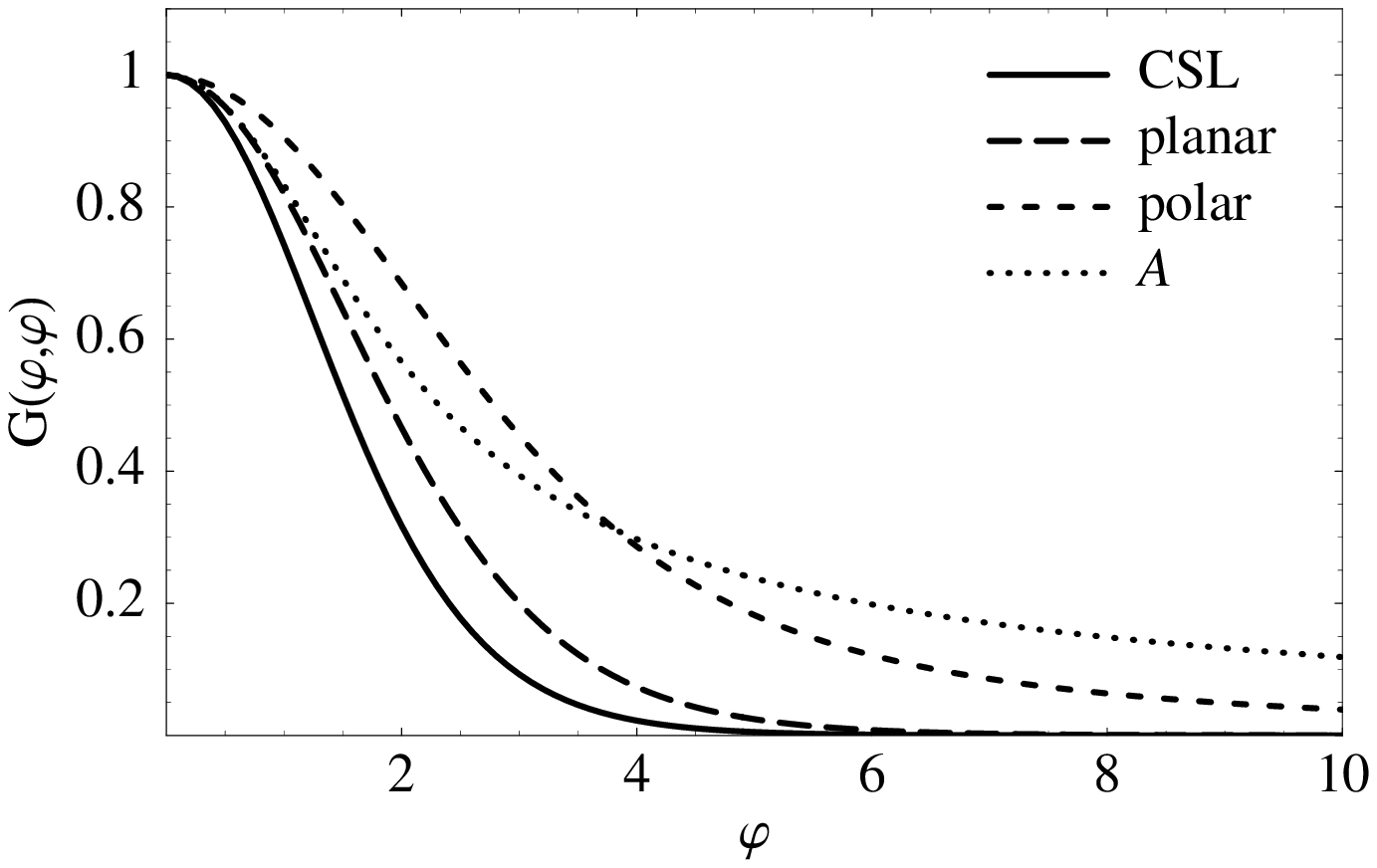}
\includegraphics[width=0.49\textwidth]{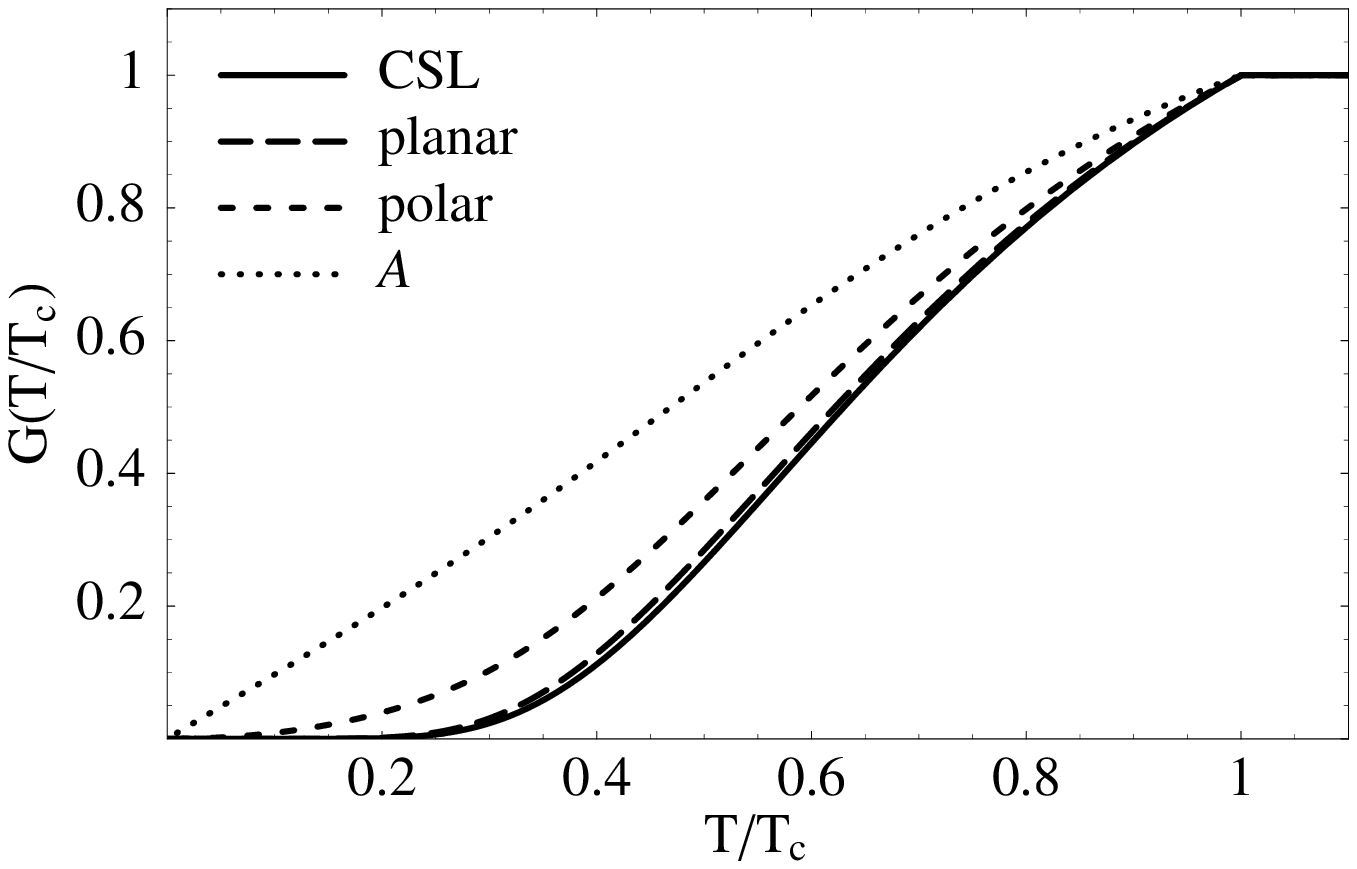}}
\vspace{0.5cm}
\caption{Left panel: the suppression functions $G(\varphi,\varphi)$ 
of the neutrino emission contributions due to gapped modes in the CSL, 
planar, polar and {\it A} phases. Right panel: the temperature dependence
of the suppression functions using Eq.~(\ref{Phi_of_T}).}
\label{figG}
\end{center}
\end{figure}

In this section, we calculate the neutrino emissivity,
\be \label{definition}
\e_\nu \equiv -\frac{\partial}{\partial t}\int\frac{d^3{\bf p}_\nu}{(2\pi)^3}\,p_\nu\,
[f_\nu (t, {\bf p}_\nu) + f_{\overline{\nu}} (t, {\bf p}_\nu)] 
= -2\frac{\partial}{\partial t}\int\frac{d^3{\bf p}_\nu}{(2\pi)^3}\,p_\nu\,
f_\nu (t, {\bf p}_\nu) \, .
\ee
This is the total energy loss per unit time and unit volume carried away from 
quark matter by neutrinos and antineutrinos. Inserting Eq.~(\ref{resultgen}) into 
Eq.~(\ref{definition}) and making use of the integral
\be\label{normalint}
\sum_{e_1,e_2=\pm}\int_0^\infty dv\,v^3 
\int_0^\infty dx \int_0^\infty dy
\left(e^{-e_1y}+ 1\right)^{-1}
\left(e^{e_2x} + 1\right)^{-1} 
\left(e^{v+e_1y - e_2x} + 1\right)^{-1} = \frac{457}{5040}\,\pi^6 \,\, ,
\ee
we obtain 
\be \label{emissivity}
\e_\nu = \frac{457}{630}\alpha_sG_F^2 T^6\m_e\m_u\m_d\,\left[
\frac{1}{3} + \frac{2}{3}\,G(\varphi_u,\varphi_d)\right] \,\, ,  
\ee
where
\begin{subequations}
\label{G-definition}
\bea
\label{G-definition1}
G(\varphi_u,\varphi_d) &\equiv& \frac{1260}{457\pi^6}\,
\int_0^\infty dv\, v^3 \,\int_{-1}^1 d\xi 
\,F_{\varphi_u\varphi_d}^{11}(\xi,v)\, \qquad (\mbox{CSL, planar, polar})\, ,\\
G(\varphi_u,\varphi_d) &\equiv& \frac{1260}{457\pi^6}\,
\int_0^\infty dv\, v^3 \,\int_{-1}^1 d\xi 
\,\left[F_{\varphi_u\varphi_d}^{11}(\xi,v)+F_{\varphi_u\varphi_d}^{22}(\xi,v)\right]
\quad (A).
\label{G-definition2}
\eea
\end{subequations}
In all phases we consider, the emissivity $\e_\nu$ consists of two contributions.  
The first contribution is given by the term 1/3 in the square brackets on the right-hand 
side of Eq.~(\ref{emissivity}). It originates from the ungapped modes: $r=2$ in the CSL, 
planar, and polar phases, and $r=3$ in the {\it A} phase. The second contribution 
is given by the term proportional to $G(\varphi_u,\varphi_d)$. 
It originates from the gapped modes.
The function $G(\varphi_u,\varphi_d)$ has to be evaluated numerically for each phase separately. 
For the sake of simplicity, we set $\varphi_u=\varphi_d\equiv\varphi$ in the following. The results 
for $G(\varphi,\varphi)$ are plotted in the left panel of Fig.~\ref{figG}. The right panel of 
Fig.~\ref{figG} shows the function $G(T)$, obtained from $G(\varphi,\varphi)$ by making 
use of the following model temperature dependence of the gap parameter, 
\be
\phi(T)=\phi_{0}\,\sqrt{1-\left(\frac{T}{T_c}\right)^2},
\label{Phi_of_T}
\ee 
with $\phi_{0}$ being the value of the gap parameter at $T=0$, and $T_c$ 
being the value of the critical temperature.

As a consistency check, we first read off from the figure that the general result in Eqs.~(\ref{emissivity}) 
and (\ref{G-definition}) reproduces the well-known expression for the neutrino emissivity 
in the normal phase. This is obtained by taking the limit $\varphi\to 0$. Of course, the 
result in this limit is the same for all considered phases. Since $G(0,0)=1$, see left panel
of the figure, we recover Iwamoto's result \cite{iwamoto}. 

In the spin-one phases, the function $G(\varphi,\varphi)$ 
describes the suppression of the emissivity due to the presence of the gap in the 
quasiparticle spectrum.  
In order to discuss this suppression for the different phases, we derive analytic results for
the asymptotic behavior of $G(\varphi,\varphi)$ for $\varphi \to \infty$. Physically, this
corresponds to the low temperature behavior. The details of the calculation are presented 
in Appendix~\ref{app:largephi}. The results are 
\be
G(\varphi,\varphi) \sim 
\left\{\begin{array}{lll}
\varphi \,e^{-\sqrt{2}\varphi} & 
\quad & \mbox{(CSL)}\, , \\ \\
\sqrt{\varphi}\,  e^{-\varphi} & 
\quad & \mbox{(planar)}\, , \\ \\
\varphi^{-2} & 
\quad & \mbox{(polar)}\, ,\\ \\
\varphi^{-1} & 
\quad & \mbox{({\it A})}\, .
\end{array}
\right.\label{eq:largephi}
\ee
The strongest suppression happens in the CSL phase, in which the gap is isotropic. 
At large values of $\varphi$, the emissivity is exponentially suppressed, which is universal and 
qualitatively the same as, for example, in a spin-zero color superconductor \cite{jaikumar}
or in superfluid neutron and/or proton matter \cite{Yak,voskresensky2}. 
{From} a physical viewpoint, this reflects the 
fact that, in a superconductor, the neutrino emission is proportional to the density 
of thermally excited quasiparticles. It is worth emphasizing, however, that the 
function $G(\varphi,\varphi)$ cannot be approximated well by the exponential 
function at small $\varphi$. For $\varphi\alt 1$, 
the actual suppression is much weaker. This is also obvious from 
the right panel of Fig.~\ref{figG}. This panel shows that, in all phases, 
the function $G(T/T_c)$ behaves almost linearly all the way down to 
$T/T_c\simeq 0.4$. In the CSL and the planar phases, the exponential suppression
starts to show up only below this point. 

The contributions from the gapped modes in the other spin-1 phases differ considerably
from the CSL result. All of them have some degree of anisotropy in the gap function. 
The second strongest suppression is seen in the planar phase. In this case, while the
gap function is anisotropic, it has no zeros. The dominant contribution comes from a
stripe around the equator of the Fermi sphere, where the gap function, which is proportional
to $\sqrt{1+\cos^2\theta_{f}}$, takes its minimum. 

In the polar phase, the gap function has point nodes at the north and south poles
of the Fermi sphere, i.e., it costs no energy to excite quasiparticles around these 
points. Therefore, these quasiparticles give the dominant contribution to the 
emissivity. For large values of the dimensionless variable $\varphi$, in particular, 
one has a power-law instead of the exponential suppression. 

The gap function in the {\it A} phase has also nodes at the north and south poles
of the Fermi sphere. However, there is a difference compared to the polar phase in 
the behavior of the dispersion relations at small angles $\theta_{f}$. 
While it is linear in $\theta_{f}$ in the polar phase, it is 
quadratic in $\theta_{f}$ in the {\it A} phase. This difference gives rise to a
different suppression of the emissivity, see Fig.~\ref{figG} and Eq.\ (\ref{eq:largephi}).

The results of this section will be used in Sec.~\ref{cooling} in order to discuss the  
effect of spin-1 superconductivity on the cooling rates of compact stars. Besides the 
neutrino emissivity, this requires the calculation of the specific heat.

\section{Specific heat} 
\label{specificheat}

\begin{figure}[t]
\hbox{\includegraphics[width=0.45\textwidth]{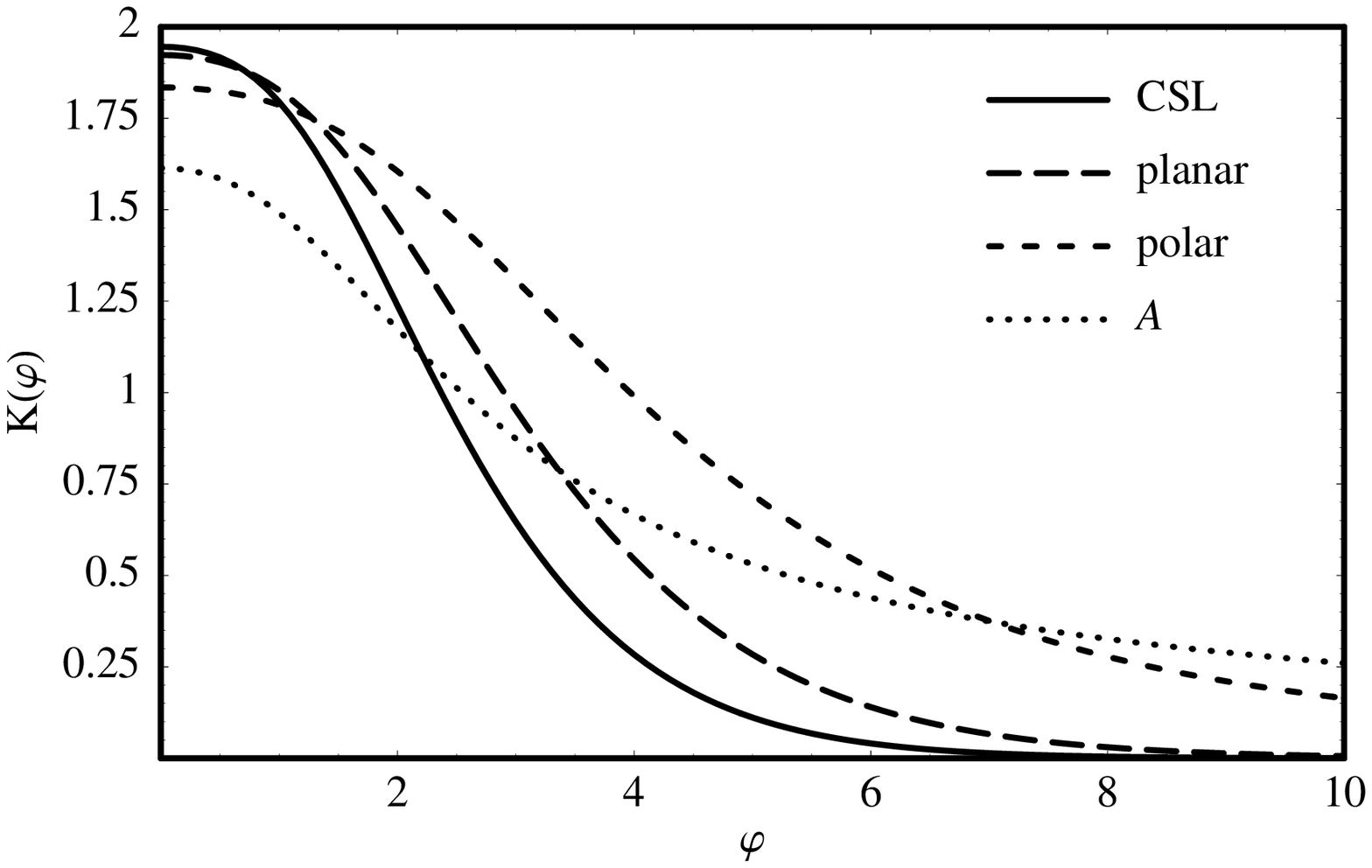}
\includegraphics[width=0.45\textwidth]{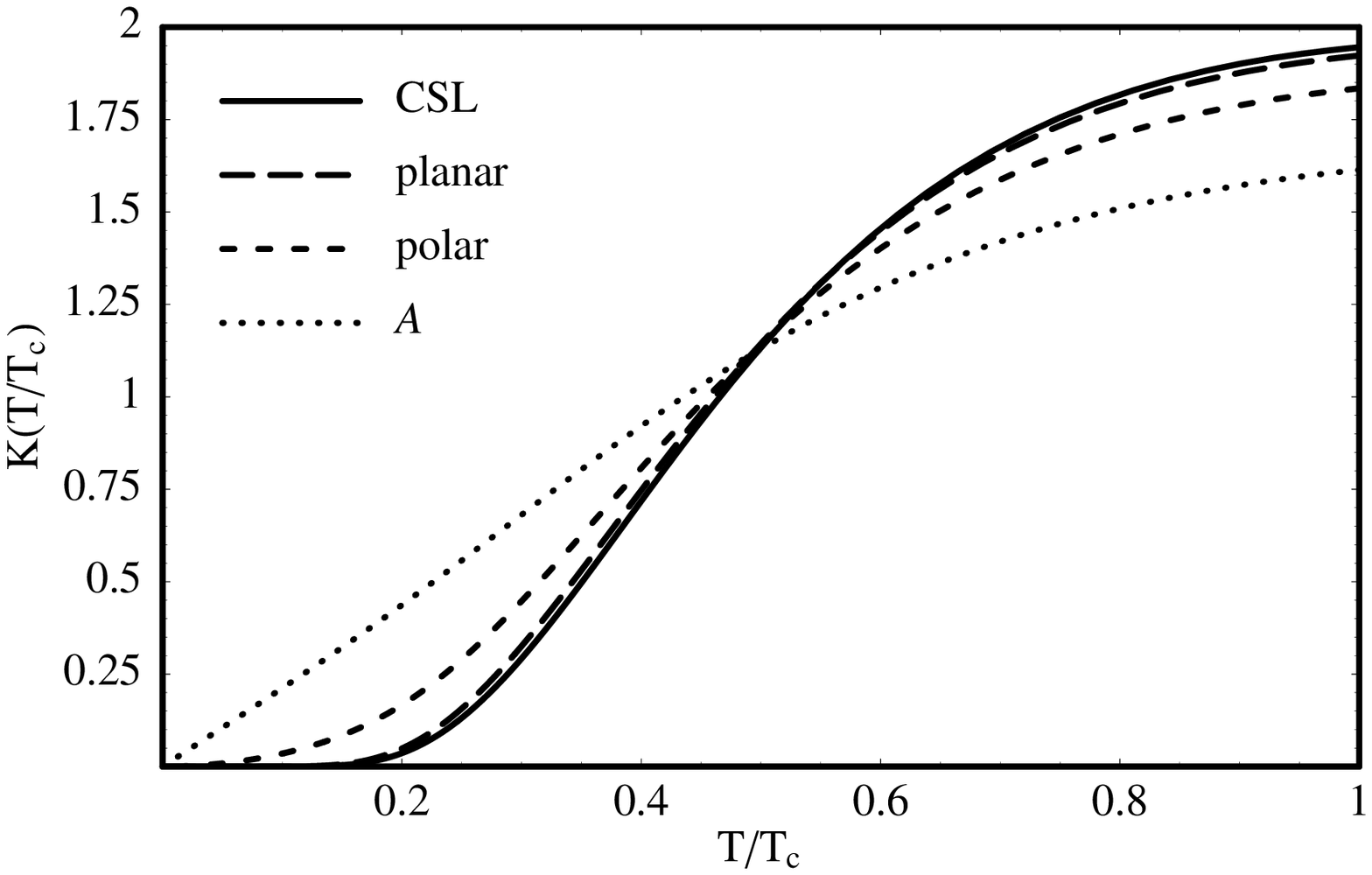}}
\caption{\label{figK}
The functions $K(\varphi)$ (left panel) 
and $K(T/T_c)$ (right panel) for four spin-one color superconductors.}
\end{figure}

In this section, we calculate the specific heat of spin-one color superconductors.
This result shall be used in the discussion of the cooling rate in the next section. 
We may start from the entropy density (see for example Ref.~\cite{vollhardt})
\be
S = -\sum_{f=u,d}\sum_{r}\frac{n_r}{2} 
\int\frac{d^3{\bf k}}{(2\pi)^3}\,\Big\{n_F(\e_{{\bf k},r,f})\,\ln n_F(\e_{{\bf k},r,f})
+ [1-n_F(\e_{{\bf k},r,f})]\,\ln[1-n_F(\e_{{\bf k},r,f})]\Big\} \,\, ,
\ee
where $n_r$ is the degeneracy of the quasiparticle branch $r$ as given in Table \ref{tablephases}.
In each phase, $\sum_r n_r/2 = 6$, accounting for 2 spin and 3 color degrees of freedom. 
The specific heat is then obtained as 
\bea \label{heatdefine}
c_V &=& T\,\frac{\partial S}{\partial T} = 
\sum_{f=u,d}\sum_{r}\frac{n_r}{2} \int\frac{d^3{\bf k}}{(2\pi)^3}
\e_{{\bf k},r,f} \frac{\partial}{\partial T}n_F(\e_{{\bf k},r,f}) 
\eea
Making use of the model temperature dependence for the gap parameter in 
Eq.~(\ref{Phi_of_T}), the result for all phases we consider can be written as
\be
c_V =  T\sum_{f=u,d} \mu_f^2 \, \left[\frac 13+\frac 23 K (\varphi_f )\right] \,\, .
\label{s-heat}
\ee
The structure of this result is analogous to that of emissivity in 
Eq.~(\ref{emissivity}), i.e., the first and the second terms in 
the square brackets on the right-hand side come from ungapped and gapped modes, respectively.
The explicit form of the function $K(\varphi)$ reads
\begin{subequations}
\bea
K(\varphi) &=& \frac{3}{\pi^2} 
\int_0^{\infty} dx \int_{-1}^{1} d\xi
\frac{e^{\sqrt{x^2+\lambda_{\xi,1}\varphi^2}}}
{\left(e^{\sqrt{x^2+\lambda_{\xi,1}\varphi^2}}+1\right)^2}
\left[x^2+\lambda_{\xi,1}\left(\varphi^2+\frac{\phi_0^2}{T_c^2}\right)\right]\, 
\qquad \mbox{(CSL, planar, polar)} ,\\
K(\varphi) &=& \frac{3}{2\pi^2} \sum_{r=1}^{2}
\int_0^{\infty} dx \int_{-1}^{1} d\xi
\frac{e^{\sqrt{x^2+\lambda_{\xi,r}\varphi^2}}}
{\left(e^{\sqrt{x^2+\lambda_{\xi,r}\varphi^2}}+1\right)^2}
\left[x^2+\lambda_{\xi,r}\left(\varphi^2+\frac{\phi_0^2}{T_c^2}\right)\right]\, 
\qquad (A) \,\, . 
\eea
\label{def-Ks}
\end{subequations}
The numerical results for the function $K(\varphi)$ for all considered 
cases are plotted in the left panel of Fig.~\ref{figK}. The 
value of this function in the limit $\varphi\to 0$ does not reproduce 
the result for the normal phase $K_{n}=1$, we rather observe  
$\lim_{\varphi\to 0}K(\varphi)>1$. This is due to the jump of the 
specific heat $\Delta c_V$ at the point of the second order phase 
transition to superconducting matter. In the model used, the magnitude 
of the jump is obtained by calculating the term proportional to 
$\phi_0^2/T_c^2$ in Eqs.~(\ref{def-Ks}). The result can be written 
as
\begin{subequations}
\bea
\Delta c_V  &=&  (\mu_u^2 + \mu_d^2) \,
\frac{2\,\overline{\phi}_1^2}{T_c\,\pi^2} 
\qquad \mbox{(CSL, planar, polar)} \,\, , \\
\Delta c_V  &=&  (\mu_u^2 + \mu_d^2) \,
\frac{\overline{\phi}_1^2+\overline{\phi}_2^2}{T_c\,\pi^2} 
\qquad (A) \,\, , 
\eea
\end{subequations}
where $\overline{\phi}_r\equiv(\int d\Omega_{\bf k}/(4\pi)\,
\lambda_{{\bf k},r})^{1/2}\,\phi_0$ is the quadratic mean of 
the $r$-th gap function. Using the results from Ref.~\cite{andreas}, 
cf. Eq.~(118) therein, we conclude that the jump of the specific 
heat is proportional to the condensation energy (at $T=0$).
Hence the order of the values $K(0)$ in Fig.~\ref{figK} 
reflects the order of the condensation energies. 

As for the emissivities, we derive 
analytical approximate expressions for the specific heat at asymptotically 
large $\varphi$, corresponding to asymptotically small temperatures. For the details of the 
calculation see Appendix~\ref{cV}. We find
\be \label{Kasym}
K(\varphi) \sim \left\{\begin{array}{lll}
\varphi^{5/2}\, e^{-\sqrt{2}\varphi} & 
\quad & \mbox{(CSL)} \, ,\\ \\
\varphi^{2}\, e^{-\varphi} & 
\quad & \mbox{(planar)}  \, ,\\ \\
\varphi^{-2} & 
\quad & \mbox{(polar)} \, ,\\ \\
\varphi^{-1} & 
\quad & \mbox{({\it A})} \, .
\end{array} 
\right. 
\ee 
This behavior of the specific heat leads to the fact that the 
curves in Fig.~\ref{figK} reverse their order at large $\varphi$ 
compared to $\varphi=0$, e.g., the specific heat in the CSL phase, 
which has the largest jump at the critical temperature, exhibits 
the largest suppression for very small temperatures.

Note that the degree of the suppression due to gapped modes 
in the specific heat and in the emissivity are similar at 
large $\varphi$, cf.\ Eqs.\ (\ref{eq:largephi}) and (\ref{Kasym}).

\section{Cooling rates} 
\label{cooling}

In this section, we shall use the results for the emissivity in Eq.~(\ref{emissivity})
the specific heat in Eq.~(\ref{s-heat}) in order to study cooling of bulk matter in 
spin-one color-superconducting phases. When the cooling is only due to the neutrino 
emissivity, one has the following relation,
\be
\epsilon_{\nu}(T) = -c_V(T) \frac{dT}{dt}\, .
\label{e=dT/dt}
\ee
In order to derive the change of temperature in time, one has to integrate the above 
equation,
\be
t-t_0 = -\int_{T_0}^{T}dT^{\prime}\frac{c_V(T^\prime)}{\epsilon_{\nu}(T^\prime)} \, ,
\label{t-t0}
\ee
where $T_0$ is the temperature at time $t_0$. By inserting the expressions
from Eqs.~(\ref{emissivity}) and (\ref{s-heat}) into Eq.~(\ref{t-t0}) and using 
$\varphi_u=\varphi_d\equiv\varphi$, we derive 
\be
t-t_0 = -\frac{630}{457} \frac{\mu_u^2+\mu_d^2}{\alpha_s G_F^2 \mu_e \mu_u \mu_d}
\int_{T_0}^{T}\frac{dT^{\prime}}{(T^{\prime})^5} 
\frac{1+2K(T^{\prime})}{1+2G(T^{\prime})}\, ,
\label{timetime}
\ee
where the temperature-dependent functions $K(T)$ and $G(T)$ are obtained 
from the functions $K(\varphi)$ and $G(\varphi,\varphi)$ with the help of 
Eq.~(\ref{Phi_of_T}). 

By making use of Eq.~(\ref{timetime}), let us estimate the cooling behavior 
of a compact star whose core is made out of spin-one color-superconducting quark matter. 
We start from the moment when the stellar core, to a good approximation, 
becomes isothermal. At this point, the stellar age is 
of the order of $t_0=10^2$~yr
and the temperature is about $T_0=100$~keV. The estimates in the literature 
\cite{ren,Schmitt:2002sc} suggest that the value of the critical temperature 
in spin-one color superconductors is of the order of $T_c=50$~keV. This is 
the value that we use in the numerical analysis. 
Moreover, we choose $\m _u=400$ MeV, $\m _d=500$ MeV, $\m _e=100$ MeV, $\a _s=1$. 
The Fermi 
weak coupling constant is given by $G_F=1.16637 \times 10^{-11}$ MeV$^{-2}$. 

The numerical results show that the cooling behavior is dominated by the ungapped 
modes. Consequently, to a very good approximation, the time dependence of the
temperature can be computed by neglecting the functions $K(T^\prime)$ and 
$G(T^\prime)$ in Eq.~(\ref{timetime}). In this case, an analytical expression 
can be easily derived,
\be
T(t) = \frac{T_0\,\tau^{1/4}}{(t-t_0+\tau)^{1/4}} \,\, , 
\ee
where
\be
\tau \equiv \frac{315}{914}
\frac{\mu_u^2 + \mu_d^2}{\alpha_s\,G_F^2\,\mu_e\,\mu_u\,\mu_d}\, \frac{1}{T_0^4}
\,\, .
\ee
With the above parameters, this constant is of the order of several minutes, 
$\tau \simeq 10^{-5}$yr.

\begin{figure}[t]
\includegraphics[width=0.6\textwidth]{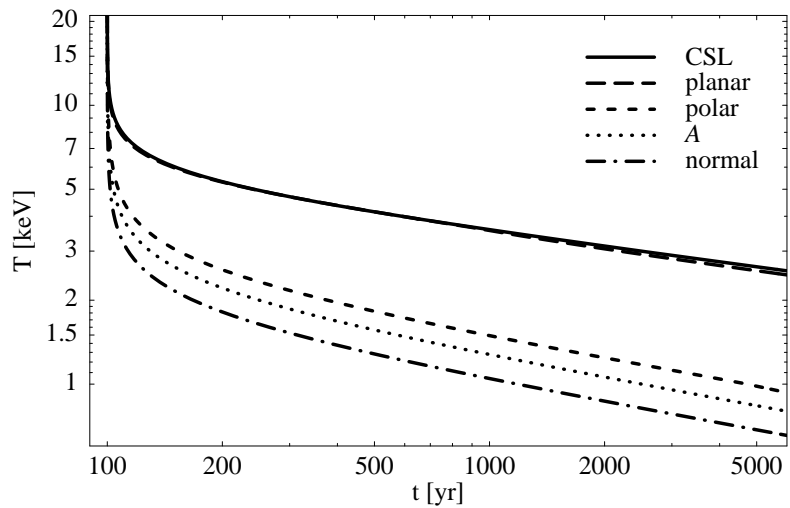}
\caption{ \label{ctime}
Temperature as a function of time for normal quark matter and four spin-one ``toy phases'' (dropping
the ungapped modes). The curves represent the CSL phase (solid), planar phase (long-dashed),
polar phase (short-dashed), {\it A} phase (dotted), and normal quark matter (dashed-dotted).}
\end{figure}

It may be interesting, although unphysical, to compare the 
cooling behavior of the gapped modes of the different 
spin-one phases. To this end, we drop the 1 in the numerator 
and denominator of the integrand in Eq.~(\ref{timetime}). 
The results are shown in Fig.~\ref{ctime}. Note that both 
the initial temperature $T_0$ and the critical temperature 
$T_c$ are beyond the scale of the figure. The reason is that, 
even for the gapped modes, the cooling time scale for temperatures
down to approximately $10$~keV is set by the above constant 
$\tau$. Therefore, all phases cool down very fast, and the 
transition to the superconducting phase at $T=50$~keV is hidden in the 
almost vertical shape of the curve. Only at temperatures 
several times smaller than $T_c$, i.e., of the order of 
$10$~keV, substantial differences between the phases appear. In this range, the fully gapped
phases cool down considerably slower than  
the phases with nodes on the Fermi sphere, which, in turn, 
cool slower than the normal phase. It seems to agree with physical intuition that this order 
reflects the order of the suppression at low temperatures for the neutrino emissivity, i.e.,   
the slowest cooling (isotropic gap) happens for the phase where  
$\epsilon_\nu$ is suppressed strongest while the fastest cooling 
(no gap) happens for the smallest suppression. Note, however, that the 
cooling depends on the ratio of the suppressions of $\epsilon_\nu$ and $c_V$. Therefore, this 
order is a nontrivial consequence of the exact forms of the functions 
$G(\varphi,\varphi)$ and $K(\varphi)$. For large values of $\varphi$, we may use 
Eqs.\ (\ref{eq:largephi}) and (\ref{Kasym}) to estimate the ratio $K(\varphi)/G(\varphi,\varphi)$.
For both completely gapped phases we find $K(\varphi)/G(\varphi,\varphi) \sim \varphi^{3/2}$ while 
both phases with point nodes yield ratios independent of $\varphi$. Consequently, for late
times, $T\sim t^{-2/11}$ in the CSL and planar phases while $T\sim t^{-1/4}$ in the
polar, {\it A} and normal phases. 


\section{Spatial asymmetry in the neutrino emission from the {\it A} phase}
\label{kicks}

In this section, we address a special aspect 
of the angular distribution of the neutrino emission. To this end, we consider the 
net momentum carried away by neutrinos and antineutrinos from the quark system per unit volume and time, 
\bea \label{H-definition}
\frac{d{\bf P}^{(net)}}{dV \, dt}
\equiv -\frac{\partial}{\partial t}\int\frac{d^3{\bf p}_\nu}{(2\pi)^3}\,{\bf p}_\nu\,
[f_\nu (t, {\bf p}_\nu) +f_{\overline{\nu}} (t, {\bf p}_\nu)]  
= -2\frac{\partial}{\partial t}\int\frac{d^3{\bf p}_\nu}{(2\pi)^3}\,{\bf p}_\nu\,
f_\nu (t, {\bf p}_\nu) \,.
\eea
Analogously to the case of the total emissivity, see Eq.~(\ref{emissivity}), we arrive 
at the following general result,
\be \label{P-emissivity}
\frac{d{\bf P}^{(net)}}{dV \, dt} 
= \frac{457}{945}\alpha_sG_F^2 T^6\m_e\m_u\m_d\,H(\varphi_u,\varphi_d)\, \hat{\bf z} \,\, ,  
\ee
where $\hat{\bf z}$ is the unit vector in $z$ direction, and 
\begin{subequations}
\label{H-definition0}
\bea
\label{H-definition1}
H(\varphi_u,\varphi_d) &\equiv& -\frac{420}{457\pi^6}\,
\int_0^\infty dv\, v^3 \,\int_{-1}^1 d\xi \xi 
\,F_{\varphi_u\varphi_d}^{11}(\xi,v)=0\, \qquad (\mbox{CSL, planar, polar})\, ,\\
H(\varphi_u,\varphi_d) &\equiv& -\frac{420}{457\pi^6}\,
\int_0^\infty dv\, v^3 \,\int_{-1}^1 d\xi \xi 
\,\left[F_{\varphi_u\varphi_d}^{11}(\xi,v)+F_{\varphi_u\varphi_d}^{22}(\xi,v)\right]
\quad (A).
\label{H-definition2}
\eea
\end{subequations}
In the CSL, planar, and polar phases, the function $H(\varphi_u,\varphi_d)$ 
is identically zero. This is because $F_{\varphi_u\varphi_d}^{11}(\xi,v)$ is 
an even function of $\xi$ in these three phases, and therefore the integration 
over $\xi$ in Eq.~(\ref{H-definition1}) is vanishing. This means that the 
net momentum of emitted neutrinos as well as the related net recoil momentum 
of bulk quark matter in the CSL, planar, and polar phases are zero. 

\begin{figure}[t]
\begin{center}
\hbox{\includegraphics[width=0.49\textwidth]{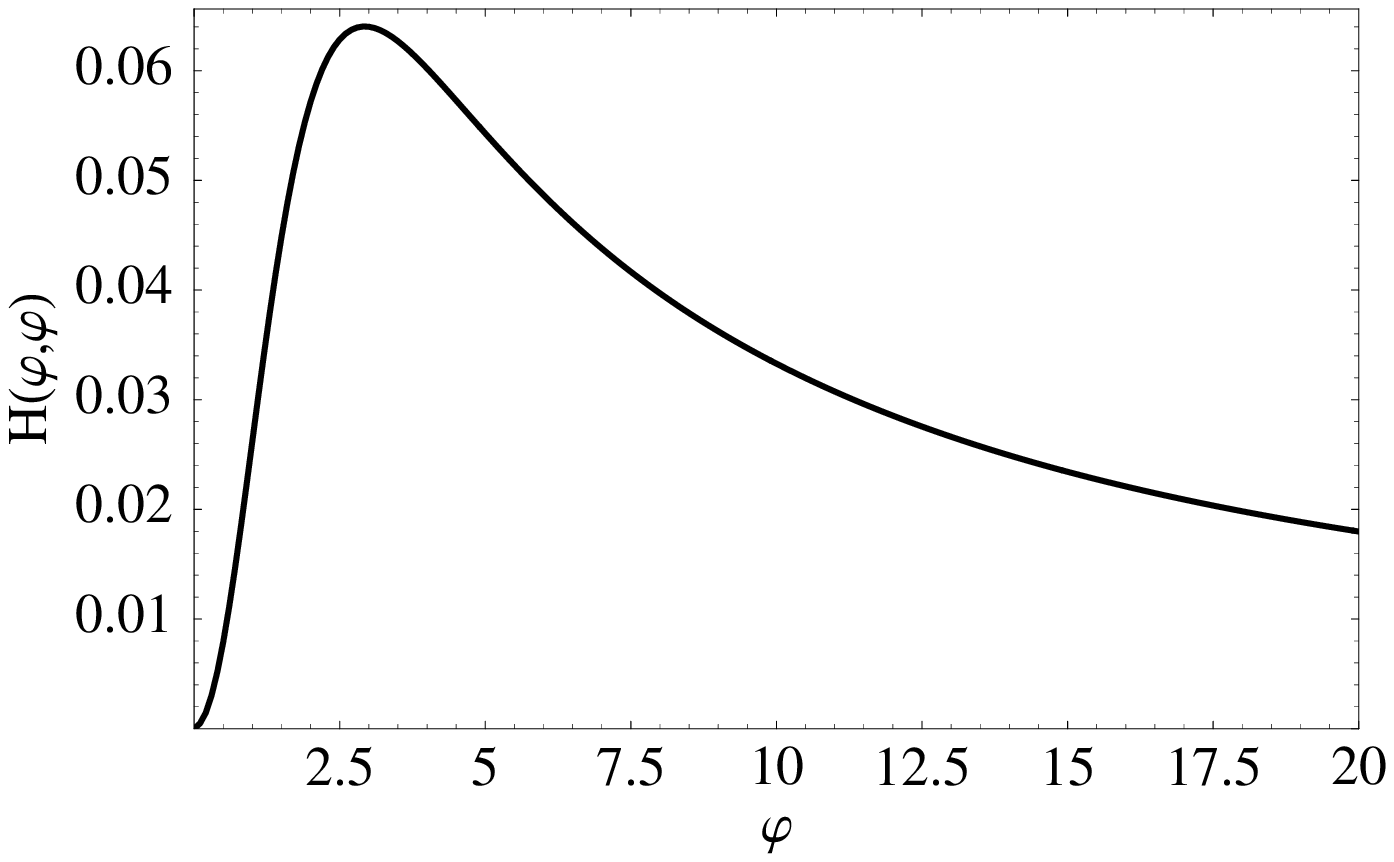}
\includegraphics[width=0.49\textwidth]{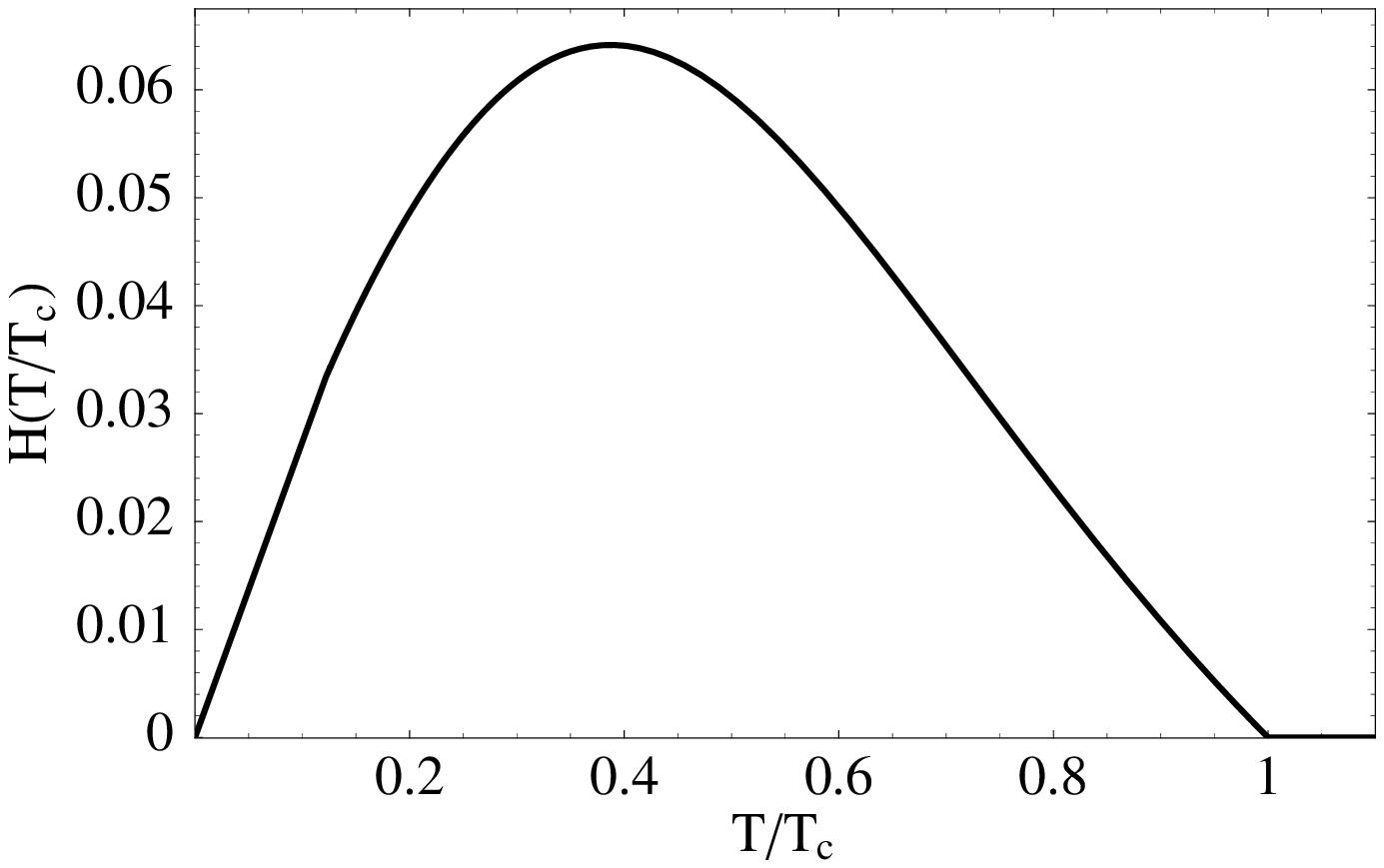}}
\caption{Numerical results for the functions $H(\varphi,\varphi)$ and $H(T/T_c)$
which determine the net momentum carried away from the spin-1 color superconducting 
{\it A} phase by neutrinos.}
\label{figH}
\end{center}
\end{figure}

The result is non-vanishing, however, in the {\it A} phase. The 
corresponding function $H(\varphi,\varphi)$ is plotted in the left panel 
of Fig.~\ref{figH}. From the figure, we see that $H(0,0)=0$. 
Of course, this is just a consistency check that, in the limit 
$\varphi\to 0$, we reproduce the vanishing result in the fully 
isotropic normal phase of quark matter. From the numerical data, 
we find that the maximum value of the function $H(\varphi,\varphi)$ 
is approximately equal to $0.064$, which corresponds to the value 
of its argument $\varphi\simeq 2.9$. At large $\varphi$, the 
asymptotic behavior of $H(\varphi,\varphi)$ is power suppressed 
as $1/\varphi$. For completeness, we also show the 
temperature-dependent function $H(T)$ in the right panel of 
Fig.~\ref{figH}. This is obtained from $H(\varphi,\varphi)$ by 
making use of the model temperature dependence for the gap in 
Eq.~(\ref{Phi_of_T}).

It may look surprising that the net momentum from the {\it A} phase is nonzero,
indicating an asymmetry in the neutrino emission with respect to the reflection
of the $z$-axis. The gap functions do not exhibit this asymmetry. 

The origin of this remarkable result can be made transparent by rewriting the 
expression (\ref{H-definition2}) for the {\it A} phase in the following way,
\be
H\left(\varphi_u,\varphi_d\right)
= -\frac{840}{457\pi^6} \int_0^\infty d v v^3 \int_{-1}^1 d \xi \, \xi \, 
F^{\rm eff}_{\varphi_u \varphi_d}(\xi,v)\, ,
\label{H-function}
\ee
where
\bea
F^{\rm eff}_{\varphi_u \varphi_d}(\xi,v) &\equiv& 
\sum_{e_1,e_2=\pm}
\int_0^\infty \int_0^\infty 
\,dx\,dy
\left(e^{-e_1\sqrt{y^2+(1+\xi)^2\varphi_u^2}}+ 1\right)^{-1}
\left(e^{e_2\sqrt{x^2+(1+\xi)^2\varphi_d^2}} + 1\right)^{-1} \non 
&&\times\; \left(e^{v+e_1\sqrt{y^2+(1+\xi)^2\varphi_u^2} - 
e_2\sqrt{x^2+(1+\xi)^2\varphi_d^2}} + 1\right)^{-1} \, .
\label{F-eff}
\eea
In the derivation, we used the explicit forms of $\omega_{rr}(\xi)$ and $\lambda_{\xi,r}$
from Table~\ref{tableomega}. Now, the result looks as if only one single quasiparticle 
mode contributes to the net neutrino momentum. The corresponding ``effective'' gap function 
has the angular dependence $\sim (1 + \xi)$ which clearly discriminates between $+z$ and $-z$ 
directions. 

In order to understand the physical reason for the appearance of the effective quasiparticle 
mode, it is useful to analyze the physical properties of the gapped modes of the {\it A}
phase, $r=1,2$. The color-spin structure of these modes is encoded in the projection 
operators ${\cal P}^{+}_{{\bf k},r}$. Their explicit form is given in Eq.~(\ref{proj-P_A}).
It is instructive to write the first two projectors in the form
\begin{subequations}
\label{P12+}
\bea
\label{P1+}
{\cal P}^{+}_{{\bf k},1} &=& \frac{1}{2}J_{3}^2[1-{\rm sgn}(\hk_3)]\,H^+(\uk) +
\frac{1}{2}J_{3}^2[1+{\rm sgn}(\hk_3)]\,H^-(\uk) \,\, ,\\
{\cal P}^{+}_{{\bf k},2} &=& \frac{1}{2}J_{3}^2[1+{\rm sgn}(\hk_3)]\,H^+(\uk) +
\frac{1}{2}J_{3}^2[1-{\rm sgn}(\hk_3)]\,H^-(\uk) \,\, ,
\label{P2+}
\eea
\end{subequations}
where $H^\pm(\uk) \equiv \frac{1}{2}(1\pm \vS\cdot\uk)$ are 
the helicity projectors with $\vS\equiv\g^5\g^0\vg$. From 
Eq.~(\ref{P1+}) we see that the quasiparticles of the first 
branch have helicity $+1$ when the projection of their momentum 
onto the $z$-axis is negative, $\hk_3<0$, and helicity $-1$ 
if $\hk_3>0$. Quasiparticles of the second branch have opposite 
helicities, see Eq.~(\ref{P2+}).

\begin{figure}[t]
\includegraphics[width=0.45\textwidth]{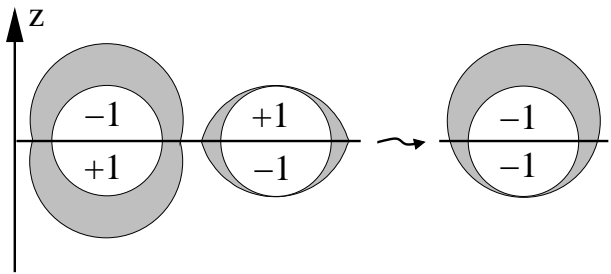}
\caption{\label{effgap}
Gap functions for the first (left) and the second (middle) excitation
branch with specified helicities of quasiparticles in the upper and
the lower hemispheres. The ``effective'' gap relevant for the neutrino
emission is shown on the right.}
\end{figure}

The next step in the argument is to notice that only left-handed quarks 
participate in the weak interactions which underly the Urca processes. 
Formally, this can be seen from Eq.~(\ref{Pi1}) where the left 
chirality projectors $\frac12(1-\g^5)$ occur in the first term 
under the trace. (Note that the second term describes charge-conjugate 
quarks for which $\frac12(1+\g^5)$ projects also onto 
left chirality states.) In the ultrarelativistic limit, these 
are quarks with negative helicity. Taking into account the 
helicity properties of the quasiparticles in the {\it A} phase, 
it becomes clear that only an effective gap structure contributes. 
This is constructed from the upper hemisphere of the first mode 
and the lower hemisphere of the second mode, see Fig.~\ref{effgap}.
This is a graphical representation of the formal argument given 
after Eq.~(\ref{F-eff}). [Of course, our choice for the angular 
dependence of the gap functions, namely
$\lambda_{{\bf k},1} = (1+|\cos\theta_{\bf k}|)^2$ and 
$\lambda_{{\bf k},2} = (1-|\cos\theta_{\bf k}|)^2$, is 
only one possible convention. Equivalently, one could choose 
$\lambda_{{\bf k},1} = (1+\cos\theta_{\bf k})^2$ and 
$\lambda_{{\bf k},2} = (1-\cos\theta_{\bf k})^2$, in which 
case the quasiparticle excitations would be ordered according 
to their helicity. Then, quasiparticles of the first (second) 
branch would have negative (positive) helicity, and the weak 
interaction would involve only quasiparticles of the first 
branch. Our convention in this paper is in accordance with 
Ref.~\cite{andreas}.]
 
The asymmetry in the effective gap function translates into the 
asymmetry of the neutrino emission. This is due to the  
angular dependence of the amplitude for Urca type processes. As in 
the vacuum, the corresponding amplitude is proportional to $1-\cos
\theta_{\nu d}$. This can be seen, for instance, from the integrand 
in Eq.~(\ref{varchange}). Such an angular dependence of the amplitude
means that the neutrinos are emitted preferably in the direction 
opposite to the (almost collinear) momenta of the participating 
up and down quarks. In fact, this is a general property that 
holds also in the normal phase \cite{iwamoto}. Since the effective 
gap function assumes smaller values for quasiparticles with $\hk_3<0$ than 
with $\hk_3>0$, there is more neutrino emission in the $+z$ 
direction.

One can estimate the maximum velocity of a neutron star with 
a quark matter core in the {\it A} phase that can be obtained by the
asymmetric neutrino emission. It has been shown that this velocity
is negligibly small, e.g., of the order 1~m/s, see erratum in 
Ref.~\cite{prl}. In essence, the reason for this is that 
the available thermal energy in the star, after matter in the 
stellar interior cools down to the critical temperature 
$T_c \lesssim 100$~keV of the {\it A} phase, is too small to 
power substantial momentum kicks. (It would be interesting to 
investigate, however, if additional sources of stellar heating, 
e.g., such as the latent heat from a first-order phase transition, 
could change the conclusion.)

\section{Quark mass effects}
\label{quarkmass}

In nature, quarks are not exactly massless. Therefore, it is 
important to address the effects that the masses have on the 
dispersion relations of quasiparticles, and thus on the neutrino 
emissivity and the specific heat of quark matter. 

In order to study the massive case, we keep the color-spin structure
of the gap matrix ${\cal M}_{\bf k}$ exactly as in the massless case, 
given in Eq.~(\ref{defM}). Then, the inverse full quark propagator can be 
written as follows
\be
S^{-1}(K)=\left(\begin{array}{cc}
\left[G^{+}_0(K)\right]^{-1} & 
{\cal M}_{\bf k} \phi \\
\gamma^0 {\cal M}_{\bf k}^{\dagger} \gamma^0\phi & 
\left[G^{-}_0(K)\right]^{-1} 
\end{array}\right) \, ,
\ee
with $\left[G^{\pm}_0(K)\right]^{-1} \equiv \gamma^\mu K_{\mu}\pm\gamma^0\mu +m$,
where $m$ is the quark mass. For simplicity, we omit the flavor index $f$ in 
this section. 

We do not repeat the detailed calculations for the emissivity 
and the specific heat with this propagator. Instead we assume that the main
modification happens due to the change of the quasiparticle dispersion relations.
These relations are determined by the solution to the algebraic equation
$\mbox{det}\,S^{-1}=0$, or more explicitly,
\be
\mbox{det}\left[ (G^{-}_0)^{-1}(G^{+}_0)^{-1}
- \phi^2 (G^{-}_0)^{-1}  {\cal M}_{\bf k}   G^{-}_0 
\gamma^0 {\cal M}_{\bf k}^{\dagger} \gamma^0\right]=0\, .
\ee
We find that, in all considered spin-one phases, the results for the dispersion 
relations are essentially the same as in the massless case, except for the 
replacement $k\to \sqrt{k^2+m^2}$,
i.e.,
\be
\epsilon_{{\bf k},r}^2= \left(\sqrt{k^2+m^2}-\mu\right)^2
+\lambda_{{\bf k},r}\phi^2 \,\,  ,
\label{spectrum-mass}
\ee
with the same $\lambda_{{\bf k},r}$'s as for $m=0$, see 
Table~\ref{tablephases}. This result suggests that the emissivity and 
the specific heat do not change qualitatively after including quark 
masses. This conclusion may not be so surprising because, 
in general, non-zero quark masses are not expected to affect much 
the physical properties which are dominated by quasiparticle states 
in the vicinity of the Fermi sphere.

In view of the above ``trivial'' effect of the quark masses, it is 
appropriate to comment on the recent study in Ref.~\cite{Blaschke} 
where it is argued that there are no ungapped modes in the CSL phase 
when the quarks are massive. This may look as a contradiction to our result
(\ref{spectrum-mass}), showing that all ungapped modes, $\lambda_{{\bf k},2}=0$, 
survive after switching on the mass.
The seeming contradiction is removed, 
however, after noticing that a different choice of the gap matrix
in the CSL phase is utilized in Ref.~\cite{Blaschke}. In our 
notation, the corresponding gap matrix would be obtained by replacing 
$\gamma_{\perp,j}(\uk)$ with $\gamma_{j}$ in Eq.~(\ref{defM}).
After making such a replacement, we find that the dispersion 
relations of quasiparticles in Ref.~\cite{Blaschke} are indeed
reproduced. In particular, the low-energy dispersion relations 
for two out of total three different quasiparticles are given by 
\bea
\e_{{\bf k},1/2}^2 &\simeq & \left(\sqrt{k^2+m^2}-\mu\right)^2
+2\,\phi^2\,\left(\sqrt{1+\frac{m^2}{8\mu^2}} \pm
\frac{m}{2\sqrt{2}\mu}\right)^2\, .
\label{e3}
\eea
Note that these are obtained without the limitation of the smallness
of the quark mass $m$. The corresponding two energy gaps are thus
\be
\phi_{1/2} = \sqrt{2}\,\phi\,\left(\sqrt{1+\frac{m^2}{8\mu^2}}
\pm\frac{m}{2\sqrt{2}\mu}\right)\, .
\ee
The low-energy approximation, 
defined by $\e_{{\bf k},1/2}/\mu \sim \phi/\mu \sim |\sqrt{k^2+m^2}-\mu|/\mu \ll 1$, 
is completely sufficient for the study of 
most transport and neutrino processes in spin-one color-superconducting 
phases.

The dispersion relation of the third quasiparticle mode can be 
extracted exactly, 
\bea
\e_{{\bf k},3}^2 &=& \mu^2+k^2+m^2+\phi^2-2\sqrt{\mu^2(k^2+m^2)+k^2\phi^2}\non
&\simeq& \left(\sqrt{k^2+m^2}-\mu\right)^2 +\frac{m^2\phi^2}{\mu^2}\, .
\label{e1} 
\eea
Hence the value of the energy gap is 
\be
\phi_{3} = \frac{m \phi}{\sqrt{\mu^2+\phi^2}} \simeq \frac{m \phi}{\mu} \,\, .
\ee
For $m=0$, we recover the gaps $\phi_1=\phi_2=\sqrt{2}\,\phi$, $\phi_3=0$. 
Thus, in contrast to the massive case, both CSL gap matrices (i.e., one with 
$\gamma_{\perp,j}(\uk)$ and the other with $\gamma_{j}$ in the definition 
of ${\cal M}_{\bf k}$) give rise to the same dispersion relations in the 
ultrarelativistic case. It should be studied in the 
future which of the two physically different CSL phases at $m\neq 0$ 
has the lower free energy.

\section{Conclusions}
\label{conclusion}

In this paper, we have computed the neutrino emissivity due to 
direct Urca processes (i.e., 
$u+e^{-} \to d+\nu$ and $d \to u+e^{-} +\bar{\nu}$),
as well as the specific heat in four different spin-one 
color-superconducting phases of dense quark matter. Starting 
from the kinetic equation, we have derived a general expression 
for the neutrino emissivity. In the case of the normal 
phase of quark matter, this reduces to the well-known 
analytical result \cite{iwamoto}. The basic ingredients in 
the calculation are the quasiparticle dispersion relations, 
containing the spin-one gap functions. We have  
studied in detail the effect of an isotropic gap function 
(CSL phase) as well as of anisotropic gap functions 
(planar, polar, {\it A} phases). The numerical results 
for the emissivity and the specific heat as functions 
of the ratio of the gap parameter to the temperature, 
$\varphi=\phi/T$, have been presented.

In all four phases, also analytical expressions have been derived 
in the large $\varphi$ limit (i.e., in the limit of small 
temperatures). In particular, in the case of an isotropic 
gap function (CSL phase), the well-known exponential 
suppression of the emissivity and the specific heat is 
observed. We find that anisotropic gaps give rise to 
different asymptotes in general. For example, the phases in 
which the gap has point nodes (i.e., polar and {\it A} phases) 
show a power-law instead of an exponential suppression at 
$\varphi\to\infty$. The actual form of the power-law 
depends on the behavior of the gap function in the 
vicinity of the nodes. While a linear behavior gives 
rise to a suppression $\sim 1/\varphi^2$, a quadratic 
behavior leads to $\sim 1/\varphi$.

We have used the results for the emissivity and the specific 
heat to discuss the cooling curves of spin-one color
superconductors. Our simplified analysis, performed for 
an infinite homogeneous system, reveals several important 
qualitative features. Most importantly, the cooling rates
of all considered spin-one color superconducting phases 
differ very little from that of the normal phase. The 
reason is that, with the choice of the gap matrix as in 
Eq.~(\ref{defM}), all phases (even if the quarks are massive) 
have an ungapped quasiparticle mode which dominates cooling 
at low temperatures and, thus, makes the suppression effect 
due to gapped modes unobservable. Note, however, that this 
conclusion may change if another choice of the gap matrix, 
e.g., such as used in Ref.~\cite{Blaschke}, corresponds to 
the true ground state of quark matter.

In addition to the cooling rates, we have discussed an unusual 
property of the color-superconducting {\it A} phase, in which 
the neutrino emission is not symmetric under the reflection 
of one of the coordinate axes in position space. This is seen 
from the fact that the net momentum of emitted neutrinos is nonzero,
pointing into a direction spontaneously picked by the order parameter. 
As we have argued, the asymmetry is related to the 
helicity properties of the quasiparticles in the {\it A} phase.
A helicity order arises naturally from the structure of 
the gap matrix which is a straightforward generalization of
the {\it A} phase in $^3$He (where, however, there is no 
helicity order).

So far, we did not find any observable consequence of the 
helicity order in the {\it A} phase. The simplest possibility 
would be realized if the asymmetric neutrino emission could
result in a ``neutrino rocket'' mechanism for stars \cite{prl}. 
The estimated effect on the stellar velocity appears to be 
extremely small, however. The search for other observable 
signatures, e.g., dealing with the timing of pulsars, may 
reveal other possibilities. We could also imagine that 
observable signatures of the helicity-type order might 
be observed in completely different systems in atomic or 
condensed matter physics (e.g., trapped gases of cold fermionic 
atoms, or high-$T_c$ superconductors) if they happen to have 
a similar structure of the order parameter, see for example 
Ref.~\cite{helicity-cond-mat}. A systematic study of such a 
possibility is outside the scope of this paper, 
however.

\begin{acknowledgments}
The authors thank D.~Blaschke, M.~Buballa, D.~K.~Hong, M.~M.~Forbes, 
P.~Jaikumar, C.~Kouvaris, S.~B.~Popov, K.~Rajagopal, A.~Sedrakian, 
and D.~H.~Rischke for comments and discussions. The work of I.A.S. was supported in part
by the Virtual Institute of the Helmholtz Association under
grant No. VH-VI-041 and by Gesellschaft f\"{u}r Schwerionenforschung
(GSI), Bundesministerium f\"{u}r Bildung und Forschung (BMBF).
A.S. thanks the German Academic Exchange Service (DAAD) for
financial support. A.S. and I.A.S. thank the Center 
for Theoretical Physics at MIT for their kind hospitality. 
Q.W. is supported in part by the startup grant from University of Science
and Technology of China (USTC) in association with 'Bai Ren' project of
Chinese Academy of Sciences (CAS).

\end{acknowledgments}

\appendix 

\section{Matsubara sum} \label{matsubara}

In the calculation of the polarization tensor $\Pi_R^{\lambda\sigma}(Q)$ in 
Sec.~\ref{formalism} we use the following Matsubara sum, 
\bea
T\sum_{k_0}\frac{(k_0+\xi_1)(k_0+q_0+\xi_2)}{(k_0^2-\e_1^2)[(k_0+q_0)^2-\e_2^2]}
&=&\frac{1}{4\e_1\e_2}\left\{\left[\frac{(\e_1-\xi_1)(\e_2+\xi_2)}{q_0-\e_1-\e_2} - 
\frac{(\e_1+\xi_1)(\e_2-\xi_2)}{q_0+\e_1+\e_2}\right]\left[1-n_F(\e_1)-n_F(\e_2)\right] \right. \non
&& \left. +\; \left[\frac{(\e_1+\xi_1)(\e_2+\xi_2)}{q_0+\e_1-\e_2} - 
\frac{(\e_1-\xi_1)(\e_2-\xi_2)}{q_0-\e_1+\e_2}\right]\left[n_F(\e_1)-n_F(\e_2)\right]\right\} \,\, ,
\label{eqA1}
\eea
where $k_0$ ($q_0$) is the fermionic (bosonic) energy and $\xi_1, \xi_2, \e_1, \e_2$ are real numbers
($\e_1, \e_2 >0$). With 
\be
\label{A2}
1-n_F(\e_1)-n_F(\e_2) = \frac{n_F(\e_1)\,n_F(\e_2)}{n_B(\e_1+\e_2)} 
= -\frac{n_F(-\e_1)\,n_F(-\e_2)}{n_B(-\e_1-\e_2)}   
\ee
and 
\be
\label{A3}
n_F(\e_1)-n_F(\e_2) = \frac{n_F(-\e_1)\,n_F(\e_2)}{n_B(-\e_1+\e_2)} = -\frac{n_F(\e_1)\,
n_F(-\e_2)}{n_B(\e_1-\e_2)}
\ee
we can write the result in the following compact form,
\be \label{eqA4}
T\sum_{k_0}\frac{(k_0+\xi_1)(k_0+q_0+\xi_2)}{(k_0^2-\e_1^2)[(k_0+q_0)^2-\e_2^2]}
=-\frac{1}{4\e_1\e_2}\sum_{e_1,e_2=\pm}\frac{(\e_1-e_1\xi_1)\,(\e_2-e_2\xi_2)}{q_0-e_1\e_1+e_2\e_2}
\,\frac{n_F(-e_1\e_1)\,n_F(e_2\e_2)}{n_B(-e_1\e_1+e_2\e_2)} \,\, .
\ee
This formula is applied to Eq.~(\ref{beforematsu}).

\section{Color and Dirac traces} \label{colordirac}

In this appendix, we compute the color and Dirac traces of the tensor 
${\cal T}_{rs,+}^{\lambda\sigma}(\uk,\up)$, see Eq.~(\ref{defT}), in 
the polar, planar, {\it A}, and CSL phases. Moreover, we contract this 
tensor with the tensor $L^{\lambda\sigma}({\bf p}_e,{\bf p})$, 
see Eq.~(\ref{defL}), to obtain the functions $\omega_{rs}(\uk,\up)$, 
see Eq.~(\ref{contract}). In this appendix, for the sake
of clarity, we shall add the subscripts $c$ or $D$ at 
the symbol ``Tr'' to indicate whether the trace is taken over 
color or Dirac space. In the calculation, we encounter 
the following expressions,
\begin{subequations}
\bea
L^{00}({\bf p}_e,{\bf p}_\nu) &=& 8\,(p_e p_\nu + {\bf p}_e\cdot{\bf p_\nu}) \,\, , \\
L^{0i}({\bf p}_e,{\bf p}_\nu) &=& 8\,[p_e p_\nu^i + p_e^i p_\nu + i\,({\bf p}_e\times{\bf p_\nu})_i] \,\, , \\
L^{i0}({\bf p}_e,{\bf p}_\nu) &=& 8\,[p_e p_\nu^i + p_e^i p_\nu - i\,({\bf p}_e\times{\bf p_\nu})_i] \,\, , \\
L^{ij}({\bf p}_e,{\bf p}_\nu) &=& 8\,[\d^{ij}(p_e p_\nu - {\bf p}_e\cdot{\bf p_\nu}) + p_e^i p_\nu^j + 
p_e^j p_\nu^i -i\,\e^{ij\ell}\,(p_e p_\nu^\ell - p_e^\ell p_\nu)] \,\, , 
\eea
\end{subequations}
where $i,j=1,2,3$, and 
\begin{subequations}
\bea
{\cal T}^{00}(\uk,\up) &=& 2\,(1+\uk\cdot\up) \,\, , \\
{\cal T}^{0i}(\uk,\up) &=& 2\,[\hk^i+\hp^i+i\,(\uk\times\up)] \,\, , \\
{\cal T}^{i0}(\uk,\up) &=& 2\,[\hk^i+\hp^i-i\,(\uk\times\up)] \,\, , \\
{\cal T}^{ij}(\uk,\up) &=& 2\,[\d^{ij}(1-\uk\cdot\up)+\hk^i\hp^j+\hk^j\hp^i+i\,\e^{ij\ell}\,
(\hk^\ell - \hp^\ell)] \,\, , 
\eea
\end{subequations}
where we abbreviated
\be
{\cal T}^{\lambda\sigma}(\uk,\up) \equiv \Tr_D\left[
\gamma^\lambda(1-\gamma^5)\,\g^0\,
\,\Lambda_{\bf k}^-\,\gamma^\sigma (1-\gamma^5)\,\g^0\,\Lambda_{\bf p}^-\right]
\,\, .
\ee
Using the above results, we calculate the Lorentz contraction
\be \label{standard}
L_{\lambda\sigma}({\bf p}_e,{\bf p}_\nu)\,{\cal T}^{\lambda\sigma}(\uk,\up)
= 64\,(p_e-{\bf p}_e\cdot\uk)\,(p_\nu - {\bf p}_\nu\cdot\up) \,\, .
\ee
This is a generic result which we shall use in the following subsections 
to compute the more complicated contractions for the spin-one phases.

\subsection{Polar phase}

The polar phase is particularly simple, since the projection operators 
${\cal P}_{{\bf k},r}^+$ do not depend on the quark momentum ${\bf k}$. 
With 
\be
{\cal P}_{{\bf k},1}^+ = J_3^2 \,\, , \qquad {\cal P}_{{\bf k},2}^+ = 1 - J_3^2 \,\, ,
\ee
we immediately find after performing the color trace
\be
\omega_{11}(\uk,\up) = 2 \,\, , \qquad \omega_{22}(\uk,\up) = 1 \,\, , \qquad 
\omega_{12}(\uk,\up) = \omega_{21}(\uk,\up)=0 \,\, . 
\ee

\subsection{Planar phase}

In the planar phase, we have
\be
{\cal P}_{{\bf k},1}^+ = \frac{1}{1+\hk^2_3}\,\left[ J_1^2\,(1-\hk_1^2) + J_2^2\,(1-\hk_2^2) - 
\{J_1,J_2\}\,\hk_1\hk_2 + J_3\,\hk_3\g^0\g^5\vg\cdot\uk \right] \,\, , 
\qquad {\cal P}_{{\bf k},2}^+ = 1 - {\cal P}_{{\bf k},1}^+ \,\, ,
\ee
where $\{J_1,J_2\}$ denotes the anticommutator of $J_1$ and $J_2$.
In order to perform the color traces, we use $J_i^4=J_i^2$ $(i=1,2)$ and 
\bea
&&\Tr_c[J_i^2]=2 \quad (i=1,2,3) \,\, , \qquad 
\Tr_c[J_1^2J_2^2]=1 \,\, , \qquad 
\Tr_c[\{J_1,J_2\}^2]=2 \,\, , \nonumber \\
&&\Tr_c[J_i^2J_3]=\Tr_c[J_i^2\{J_1,J_2\}]=0 \quad (i=1,2) \,\, .
\eea
The only apparently additional Dirac trace which occurs in this phase is in fact identical to the above one,
\be \label{add1}
\Tr_D\left[\gamma^\lambda(1-\gamma^5)\,\g^5\,\vg\cdot\uk
\,\Lambda_{\bf k}^-\,\gamma^\sigma (1-\gamma^5)\,\g^5\,\vg\cdot\up \,\Lambda_{\bf p}^-\right]
={\cal T}^{\lambda\sigma}(\uk,\up)  \,\, .
\ee
Consequently, with Eq.~(\ref{standard}) we find
\be \label{omegaplanar}
\omega_{11}(\uk,\up)=\frac{1}{2}\,[3+\eta(\uk,\up)] \,\, , \qquad 
\omega_{22}(\uk,\up)=\frac{1}{2}\,[1+\eta(\uk,\up)] \,\, , \qquad 
\omega_{12}(\uk,\up)=\omega_{21}(\uk,\up)=\frac{1}{2}\,[1-\eta(\uk,\up)] \,\, , 
\ee
where
\be \label{eta}
\eta(\uk,\up)\equiv \frac{4\hk_3\hp_3 + (\hk_1\hp_1+\hk_2\hp_2)^2 -(\hk_1\hp_2-\hk_2\hp_1)^2}
{(1+\hk_3^2)(1+\hp_3^2)} \,\, .
\ee

\subsection{{\it A} phase}

In this case, there are three different quasiparticle branches. Hence, we have three projection
operators,
\be
{\cal P}_{{\bf k},1}^+ = \frac{1}{2}J_3^2\left[1 + {\rm sgn}(\hk_3)\,\g^0\g^5\vg\cdot\uk\right] \,\, , \qquad 
{\cal P}_{{\bf k},2}^+ = \frac{1}{2}J_3^2\left[1 - {\rm sgn}(\hk_3)\,\g^0\g^5\vg\cdot\uk\right] \,\, , \qquad
{\cal P}_{{\bf k},3}^+ = 1-J_3^2 \,\, . 
\label{proj-P_A}
\ee
Again, all Dirac traces can be reduced to the previous one, because of Eq.~(\ref{add1}) and
\bea 
&&\Tr_D\left[\gamma^\lambda(1-\gamma^5)\,\g^5\,\vg\cdot\uk
\,\Lambda_{\bf k}^-\,\gamma^\sigma (1-\gamma^5)\,\g^0\,\Lambda_{\bf p}^-\right]  \non
&&=\Tr_D\left[\gamma^\lambda(1-\gamma^5)\,\g^0\,
\Lambda_{\bf k}^-\,\gamma^\sigma (1-\gamma^5)\,\g^5\,\vg\cdot\up \,\Lambda_{\bf p}^-\right]
={\cal T}^{\lambda\sigma}(\uk,\up)  \,\, .
\eea
We thus find immediately, again using Eq.~(\ref{standard}),
\bea
\omega_{11}(\uk,\up)&=& \frac{1}{2}[1+{\rm sgn}(\hk_3)][1+{\rm sgn}(\hp_3)] \,\, , \qquad 
\omega_{22}(\uk,\up)= \frac{1}{2}[1-{\rm sgn}(\hk_3)][1-{\rm sgn}(\hp_3)] \,\, , \\ 
\omega_{12}(\uk,\up)&=& \frac{1}{2}[1+{\rm sgn}(\hk_3)][1-{\rm sgn}(\hp_3)] \,\, , \qquad 
\omega_{21}(\uk,\up)= \frac{1}{2}[1-{\rm sgn}(\hk_3)][1+{\rm sgn}(\hp_3)] \,\, ,  \\
\omega_{33}(\uk,\up)&=& 1 \,\, , \qquad 
\omega_{13}(\uk,\up)=\omega_{31}(\uk,\up)=\omega_{23}(\uk,\up)=\omega_{32}(\uk,\up)=0 \,\, .
\eea
All vanishing functions $\omega_{rs}(\uk,\up)$ are zero because of a vanishing color trace. This is 
exactly as in the polar phase where also all functions $\omega_{rs}(\uk,\up)$, that include one 
gapped and one ungapped branch, vanish.

\subsection{CSL phase}

In the CSL phase, 
\be
{\cal P}_{{\bf k},1}^+ = -\frac{1}{2}\left[{\bf J}\cdot\gperp(\uk)\right]^2 \,\, , \qquad 
{\cal P}_{{\bf k},2}^+ = 1 + \frac{1}{2}\left[{\bf J}\cdot\gperp(\uk)\right]^2 \,\, . 
\ee
It is convenient to write these projectors with the help of color indices $a,b=1,2,3$,
\be
\left({\cal P}_{{\bf k},1}^+\right)_{ab} = \d_{ab} + \frac{1}{2}\g_{\perp,b}(\uk)\g_{\perp,a}(\uk) \,\, , \qquad 
\left({\cal P}_{{\bf k},2}^+\right)_{ab} = -\frac{1}{2}\g_{\perp,b}(\uk)\g_{\perp,a}(\uk) \,\, . 
\ee
Using these expressions, we compute the color and Dirac traces. The only additional nontrivial 
trace that occurs can again be expressed in terms of the previous one,
\be
\sum_{a,b=1}^3\,\Tr_D\left[\gamma^\lambda(1-\gamma^5)\,\g^0\,\g_{\perp,a}(\uk)\g_{\perp,b}(\uk)
\,\Lambda_{\bf k}^-\,\gamma^\sigma (1-\gamma^5)\,\g^0\,\g_{\perp,b}(\up)\g_{\perp,a}(\up) \,
\Lambda_{\bf p}^-\right]
=(1+\uk\cdot\up)^2\,{\cal T}^{\lambda\sigma}(\uk,\up)  \,\, .
\ee 
Then, with Eq.~(\ref{standard}) we find
\bea
\omega_{11}(\uk,\up) &=& 1 + \frac{1}{4}(1+\uk\cdot\up)^2 \,\, , \qquad
\omega_{22}(\uk,\up) = \frac{1}{4}(1+\uk\cdot\up)^2 \,\, , \nonumber \\
\omega_{12}(\uk,\up) &=& \omega_{21}(\uk,\up) = 1 - \frac{1}{4}(1+\uk\cdot\up)^2 \,\, . 
\eea
Note that we have also used the following identity,
\be
\sum_{a=1}^3 \left[\g_{\perp,a}(\uk)\right]^2 = -2 \,\, .
\ee

\section{Angular Integrals} \label{angint}

Here we compute the integral
\be
I({\bf p}_\nu)\equiv \int d\Omega_{\bf p}\int d\Omega_{\bf k} (1-\cos\theta_{\nu d})\,
\d(\cos\theta_0-\cos\theta_{ud})\, F(\uk,\up) \,\, .
\ee
The result shall be applied to Eq.\ (\ref{varchange}). Thus, the function $F(\uk,\up)$
stands for $F^{rs}_{\varphi_u\varphi_d}(\uk,\up,v)$, defined in Eq.\ (\ref{Frs}), i.e., for 
notational convenience we omit all indices since only the dependence on $\uk$ and $\up$
is of relevance. For the definition of $\cos\theta_0$ see text below Eqs.~(\ref{fapprox12}). 

There is a fixed direction 
in position space, defined by the order parameter. We choose our 
frame such that the $z$-axis points into this direction. Moreover, 
for the $d\Omega_{\bf k}$ integration, we may choose the frame such 
that the three-momentum of the $d$ quark, ${\bf p}$, lies in the $xz$-plane. 
Then, we can write the unit vectors of the momenta of $u$ quark, $d$ quark, 
and neutrino as 
\be
\uk = \left(\begin{array}{c}\sin\theta_u\cos\varphi_u \\ 
\sin\theta_u\sin\varphi_u \\ 
\cos\theta_u\end{array}
\right) \,\, , \qquad 
\up = \left(\begin{array}{c}\sin\theta_d \\ 0 \\ \cos\theta_d\end{array}
\right) \,\, , \qquad 
\up_\nu = \left(\begin{array}{c}\sin\theta_\nu\cos\varphi_\nu \\ 
\sin\theta_\nu\sin\varphi_\nu \\ \cos\theta_\nu\end{array}
\right)  \,\, ,
\ee
respectively, and $\cos\theta_{ud} = \uk\cdot\up$, $\cos\theta_{\nu d} = \up_\nu\cdot\up$. (In this 
appendix, no confusion of the angle $\varphi_u$ with the ratio $\varphi_u=\phi_u/T$ is possible.) In the 
chosen frame, the most general angular dependence of the function $F$, which applies for all 
phases we consider, is
\be
F(\uk,\up) = F(\cos\theta_u,\cos\theta_d,\sin\theta_u\sin\theta_d,\cos\varphi_u) \,\, ,
\ee
The angular dependencies enter this function through the eigenvalues $\lambda_{{\bf k},r}$, 
$\lambda_{{\bf p},s}$ and the functions $\omega_{rs}(\uk,\up)$. In the CSL phase, the eigenvalues are 
constant. However, the functions $\omega_{rs}$ depend on the scalar product $\uk\cdot\up$, wherefore
the arguments $\sin\theta_u\sin\theta_d$ and $\cos\varphi_u$ enter the function $F$. 
In all other phases, $\lambda_{{\bf k},r}$ and
$\lambda_{{\bf p},s}$ are angular dependent. They depend only on $\cos\theta_u$ and $\cos\theta_d$, 
respectively. The functions
$\omega_{rs}$ are constant in the polar phase, whereas they depend on $\cos\theta_u$ and $\cos\theta_d$ in
the {\it A} phase. In the planar phase, the angular dependence of the functions $\omega_{rs}$ seems 
complicated, see Eqs.~(\ref{omegaplanar}) and (\ref{eta}). However, with $\hp_2=0$, the only additional 
argument is $\cos\varphi_u$.

Let us abbreviate $\xi_i\equiv \cos\theta_i$ ($i=u,d,0$). Then,
\bea
I({\bf p}_\nu) &=& \int d\Omega_{\bf p} \,(1-\cos\theta_{\nu d})\,\int_0^{2\pi}d\varphi_u\int_{-1}^1 d\xi_u 
\,\d\left(\xi_0-\xi_u\xi_d - \sqrt{(1-\xi_u^2)(1-\xi_d^2)}\,\cos\varphi_u\right)\non 
&& \times \; F(\xi_u,\xi_d,\sqrt{(1-\xi_u^2)(1-\xi_d^2)},\cos\varphi_u) \,\, .
\eea
Due to the $\d$-function, the integral is only nonzero if
\be
\left|\frac{\xi_0-\xi_u\xi_d}{\sqrt{(1-\xi_u^2)(1-\xi_d^2)}}\right| < 1 \,\, ,
\ee
or, equivalently, if $\xi_u^- < \xi_u < \xi_u^+$, where
\be
\xi_u^\pm \equiv \xi_0\xi_d \pm \sqrt{(1-\xi_0^2)(1-\xi_d^2)} \,\, .
\ee
Consequently,
\bea
I({\bf p}_\nu) &=& 2\,\int d\Omega_{\bf p} \,(1-\cos\theta_{\nu d})\, \int_{\xi_u^-}^{\xi_u^+} d\xi_u \, 
\frac{F\left(\xi_u,\xi_d,\sqrt{(1-\xi_u^2)(1-\xi_d^2)},\frac{\xi_0-\xi_u\xi_d}
{\sqrt{(1-\xi_u^2)(1-\xi_d^2)}}
\right)}
{\sqrt{(1-\xi_u^2)(1-\xi_d^2)-(\xi_0-\xi_u\xi_d)^2}} \,\, .
\eea
In order to perform the $\xi_u$ integral explicitly, we introduce the 
new variable $\xi_{u}^\prime = (\xi_u - \xi_0\xi_d)/\sqrt{(1-\xi_0^2)
(1-\xi_d^2)}$ with the integration range from $-1$ to $1$. In terms of 
the new variable, the integrand becomes a non-singular function of 
$\xi_0$ when $\xi_0\to 1$. Because of the kinematics of the Urca 
processes, see text after Eqs.~(\ref{beforeFermi}), the 
actual value of $\xi_0\simeq 1 - \kappa \mu_e^2/(\mu_u\mu_d)$ is 
very close to 1. Therefore, to leading order, we may set $\xi_0=1$. 
Then, the integral over $\xi_{u}^\prime$ can be performed analytically, 
and we arrive at
\be 
I({\bf p}_\nu) \simeq 2\pi \int d\Omega_{\bf p} \,(1-\cos\theta_{\nu d})
\,F(\xi_d,\xi_d,1-\xi_d^2,1) \,\, .
\ee
The arguments in the function $F$ show that the leading 
result is obtained from $\uk\simeq\up$. Applying this approximation 
to the results of the previous appendix, we see that the functions 
$\omega_{rs}$ become constant in the case of the planar and CSL phase. 
(Note, in particular, that $\eta(\uk,\uk)=1$ in the planar phase.) 

In order to perform the $d\Omega_{\bf p}$ integration, we 
have to reinstall the second component of the $d$ quark
momentum, $\hp_2 = \sin\theta_d\sin\varphi_d$, in the 
term $\cos\theta_{\nu d}$. We may perform the integral  
over $\varphi_d$ and obtain
\be 
I({\bf p}_\nu) \simeq 4\pi^2 \int_{-1}^1 d\xi (1-\xi\cos\theta_\nu)\,F(\xi,\xi,1-\xi^2,1) \,\, .
\ee

\section{Emissivity at low temperature}
\label{app:largephi}

In this appendix, we compute the $\varphi$ dependence of the function 
$G(\varphi,\varphi)$, cf. definition (\ref{G-definition}), for 
$\varphi\to\infty$. This corresponds to the behavior of the 
neutrino emissivity of the gapped modes for small temperatures.
For the phases in which the gap function has no nodes (CSL and 
planar phases), it is useful to compute the asymptotic behavior 
of the following integral separately, 
\bea
I(\varphi,\varphi)&\equiv& \sum_{e_1,e_2=\pm}
\int_0^\infty dv \, v^3 \, \int_0^\infty dx \,\int_0^\infty dy \, 
\left(e^{-e_1\sqrt{y^2+\varphi^2}} + 1\right)^{-1}\,
\left(e^{e_2\sqrt{x^2+\varphi^2}} + 1\right)^{-1}\non
&& \times \;\left(e^{v+e_1\sqrt{y^2+\varphi^2}
-e_2\sqrt{x^2+\varphi^2}} + 1\right)^{-1} \,\, .
\eea
After performing the summation over $e_1$ and $e_2$ explicitly 
and taking into account that $\exp(\sqrt{x^2+\varphi^2})\gg 1$ and 
$\exp(\sqrt{y^2+\varphi^2})\gg 1$, we arrive at the following 
expression,
\bea
\label{int-sum}
I(\varphi,\varphi)&\simeq& 
\int_0^\infty dv \, v^3 \, \int_0^\infty dx \,\int_0^\infty dy \, 
\Bigg(
\frac{1}{e^{\sqrt{x^2+\varphi^2}}+e^{v+\sqrt{y^2+\varphi^2}}}
+\frac{1}{e^{v+\sqrt{x^2+\varphi^2}}+e^{\sqrt{y^2+\varphi^2}}}\non
&& +\frac{1}{e^{v}+e^{\sqrt{x^2+\varphi^2}+\sqrt{y^2+\varphi^2}}}
+\frac{1}{e^{v+\sqrt{x^2+\varphi^2}+\sqrt{y^2+\varphi^2}}}\Bigg)\, .
\eea
Both the third and fourth terms in the integrand yield contributions 
of the order of $\exp(-2\varphi)$. For the third term we obtain
\bea
\int_0^\infty dv \, v^3 \, \int_0^\infty dx \,\int_0^\infty dy \, 
\frac{1}{e^{v}+e^{\sqrt{x^2+\varphi^2}+\sqrt{y^2+\varphi^2}}}  
&=& -6\int_0^\infty dx \,\int_0^\infty dy \, 
e^{-\sqrt{x^2+\varphi^2}-\sqrt{y^2+\varphi^2}}\non
&& \times \,\mbox{Li}_{4}(-e^{\sqrt{x^2+\varphi^2}+\sqrt{y^2+\varphi^2}}) \non
&\simeq& 2\pi \, \varphi^5 \, e^{-2\varphi} \,\, , 
\label{2phi1}
\eea
where we used the asymptotic behavior of the polylogarithm 
$\mbox{Li}_{4}(-e^{z})\simeq -\frac{1}{24}z^4$ for $z\to \infty$. The fourth term yields 
\be
\int_0^\infty dv \, v^3 \, \int_0^\infty dx \,\int_0^\infty dy \, 
\frac{1}{e^{v+\sqrt{x^2+\varphi^2}+\sqrt{y^2+\varphi^2}}}\simeq 3\pi\,\varphi\,\exp(-2\varphi)
\label{2phi2}
\ee
We neglect both contributions (\ref{2phi1}) and (\ref{2phi2}) to the integral 
$I(\varphi,\varphi)$ since they are suppressed stronger than the first two terms 
terms in Eq.\ (\ref{int-sum}). The latter yield contributions of the order of $\exp(-\varphi)$,
\bea
I(\varphi,\varphi) &\simeq& 2\,\int_0^\infty dv \, v^3 \, \int_0^\infty dx \,\int_0^\infty dy \,
\frac{e^{-\varphi}}{e^{x^2/(2\varphi)}+e^{v+y^2/(2\varphi)}}\non
&=& 4\varphi e^{-\varphi}
\int_0^\infty dv \, v^3 \, \int_0^\infty dx \,\int_0^\infty dy \, 
\frac{1}{e^{x^2}+e^{v+y^2}}\non
&\simeq& 42.55 \,\varphi \,e^{-\varphi}\, ,
\label{I3}
\eea
where the overall number is obtained by performing the numerical
integration.

\subsection{CSL phase}

Using Eqs.~(\ref{Frr}), (\ref{G-definition1}), (\ref{I3}), and $\lambda_{\xi,1}=2$ (see Table 
\ref{tableomega}) we immediately conclude
\bea
G(\varphi,\varphi) &=& \frac{5040}{457\,\pi^6}\,I(\sqrt{2}\varphi,\sqrt{2}\varphi) \non
&\simeq& 0.69\, \varphi\,e^{-\sqrt{2}\varphi} \,\, .
\eea 

\subsection{Planar phase}

For the planar phase, we may also use Eq.~(\ref{I3}) because the 
gap function has no nodes around the Fermi sphere. However, the 
result contains an additional angular integration. Using 
Eqs.~(\ref{Frr}), (\ref{G-definition1}), and $\lambda_{\xi,1}=1+\xi^2$, 
we derive
\be
G(\varphi,\varphi) = \frac{2520}{457\,\pi^6}\,\int_{-1}^1 d\xi\,
I(\sqrt{1+\xi^2}\,\varphi,\sqrt{1+\xi^2}\,\varphi) \, .
\ee
Approximating the angular integral as
\be
\int_{-1}^1 d\xi \,\sqrt{1+\xi^2}\,e^{-\sqrt{1+\xi^2}\varphi} \simeq \int_{-\infty}^\infty d\xi \,
e^{-(1+\xi^2/2)\varphi} = \sqrt{2\pi}\,\frac{e^{-\varphi}}{\sqrt{\varphi}} \,\, , 
\ee
we obtain
\be
G(\varphi,\varphi) \simeq  0.61 \, \sqrt{\varphi}\,e^{-\varphi} \,\, .
\ee

\subsection{Polar phase}

In the polar phase, the low temperature behavior of the neutrino 
emissivity is dominated by the regions around the nodes of the 
gap function. Since $\lambda_{{\bf k},1}=\sin^2 \theta_{\bf k}$, 
the dominant contribution comes from the vicinities of the north 
and south pole of the Fermi sphere. The gap behaves linear around 
these points, $\sqrt{\lambda_{{\bf k},1}} \simeq \theta_{\bf k}$. 
To obtain the leading order result in $\varphi$, we integrate
over the region where the quasiparticle energy is less than or 
of the order of the scale set by the temperature, i.e., 
$\sqrt{\lambda_{{\bf k},1}}\,\phi \lesssim \pi\, T$, or, equivalently, 
$\theta_{\bf k} \lesssim \pi/\varphi$ (and analogously for the south 
pole of the Fermi sphere). Consequently, we may use 
Eq.~(\ref{G-definition1}), set $\varphi_u=\varphi_d=0$ and integrate
over the appropriate angular regions. Then, upon using 
$\omega_{11}(\xi)=2$ (see Table \ref{tableomega}) and the integral 
(\ref{normalint}) we obtain 
\be
G(\varphi,\varphi) \simeq \int_0^{\pi/\varphi}d\theta_{\bf k}\,
\theta_{\bf k} = \frac{\pi^2}{2\varphi^2} 
\simeq \frac{4.93}{\varphi^2} \,\, .
\ee

\subsection{{\it A} phase}

As in the polar phase, the dominant contribution comes from the nodes 
of the gap function. Therefore, we can neglect the term corresponding 
to the first (fully gapped) branch in Eq.~(\ref{G-definition2}). 
We keep the term that corresponds to the gap function given by
$\lambda_{{\bf k},2}=(1-|\cos\theta_{\bf k}|)^2$. In contrast to the 
polar phase, the gap function behaves quadratically in the angular 
directions in the vicinity of the nodes, $\sqrt{\lambda_{{\bf k},2}} 
\simeq \theta_{\bf k}^2/2$. Therefore, in order to be consistent with 
the estimate in the polar phase, we restrict the integral to the regions 
$0 < \theta_{\bf k} < \sqrt{2\pi /\varphi}$ and $\pi - \sqrt{2\pi/\varphi} 
< \theta_{\bf k} < \pi$. With $\omega_{22}(\xi)= 1-{\rm sgn}(\xi)$ we obtain  
\be
G(\varphi,\varphi) \simeq \frac{1}{2}\int_{\pi-\sqrt{2\pi/\varphi}}^\pi 
d\theta_{\bf k}\,(\pi-\theta_{\bf k}) = \frac{\pi}{2\varphi} 
\simeq \frac{1.57}{\varphi} \,\, .
\ee
Note that due to the factor $\omega_{22}(\xi)$, in fact only one of the nodes, 
$\theta_{\bf k}=\pi$, yields a nonzero contribution. The physical reason for 
this is explained in detail in Sec.~\ref{kicks}.

\section{Specific heat at low temperature}
\label{cV}

Here we compute the behavior of the function $K(\varphi)$, cf. definition (\ref{def-Ks}),
for $\varphi\to\infty$. Physically, this corresponds to the behavior of the specific
heat of the gapped modes for small temperatures. From Eqs.~(\ref{def-Ks}) we conclude that
for large $\varphi$ and hence $\varphi^2 \gg \phi_0^2/T_c^2$
\begin{subequations}
\bea
K(\varphi)&\simeq& \frac{3}{2\pi^2}\int_0^\infty dx\,\int_{-1}^1d\xi \,
(x^2+\lambda_{\xi,1}\,\varphi^2)\,
\frac{1}{1+\cosh\left(\sqrt{x^2+\lambda_{\xi,1}\varphi^2}\right)}
\qquad \mbox{(CSL, planar, polar)} \,\, , \\
K(\varphi)&\simeq& \frac{3}{4\pi^2}\sum_{r=1}^2 
\int_0^\infty dx\,x^2\int_{-1}^1d\xi \,(x^2+\lambda_{\xi,r}\varphi^2)\,
\frac{1}{1+\cosh\left(\sqrt{x^2+\lambda_{\xi,1}\varphi^2}\right)}
\qquad (A) \,\, . \label{KA}
\eea
\end{subequations}

\subsection{CSL phase}

In the CSL phase, $\lambda_{\xi,1}=2$, thus the angular integration $d\xi$ is trivial. Moreover, we may
approximate
\be
\frac{1}{1+\cosh\left(\sqrt{x^2+\lambda_{\xi,1}\varphi^2}\right)}
\simeq 2\, e^{-\sqrt{x^2+2\varphi^2}} 
\simeq 2\, e^{-\sqrt{2}\,\varphi - \frac{x^2}{2\sqrt{2}\varphi}} \,\, .
\ee
Consequently,
\be
K(\varphi) \simeq \frac{6}{\pi^2}\int_0^\infty dx\,(x^2+2\varphi^2)\,
e^{-\sqrt{2}\,\varphi - \frac{x^2}{2\sqrt{2}\varphi}} 
\simeq \frac{3\sqrt{2}}{\pi^{3/2}}\,(\sqrt{2}\,\varphi)^{5/2}\,e^{-\sqrt{2}\,\varphi} 
\simeq 1.81\,\varphi^{5/2}\,e^{-\sqrt{2}} \,\, .
\ee

\subsection{Planar phase}

As in the CSL phase, the gap has no nodes in the planar phase. 
Therefore, a similar approximation can be used, 
\bea \label{Kplanar1}
K(\varphi)&\simeq& \frac{3}{\pi^2}\int_{-1}^1d\xi\, 
e^{-\sqrt{1+\xi^2}\,\varphi}\,\int_0^\infty dx\,\left[x^2
+(1+\xi^2)\,\varphi^2\right]\, e^{-\frac{x^2}{2\sqrt{1+\xi^2}\,\varphi}} \non
&\simeq&\frac{3}{\sqrt{2}\,\pi^{3/2}}\,\varphi^{5/2}\,
\int_{-1}^1 d\xi\,e^{-\sqrt{1+\xi^2}\,\varphi} \, (1+\xi^2)^{5/4} \,\, , 
\eea
where the contribution originating from the term $\sim x^2$ in the square bracket of the first 
line of Eq.~(\ref{Kplanar1}), has been neglected since it is suppressed by a power of $\varphi$. 
We can further approximate this expression to obtain
\be
K(\varphi)\simeq\frac{3}{\sqrt{2}\,\pi^{3/2}}\,\varphi^{5/2}\, \int_{-\infty}^{\infty}d\xi \, 
e^{-(1+\xi^2/2)\,\varphi} = \frac{3}{\pi}\,
\varphi^{2} \, e^{-\varphi} \simeq 0.95\,\varphi^2\,e^{-\varphi}\,\, .
\ee

\subsection{Polar phase}

In the polar phase, with $\lambda_{{\bf k},1}=\sin^2 \theta_{\bf k}$, 
we have
\be
K(\varphi) =  \frac{3}{2\pi^2}\int_0^\infty dx\,
\int_0^\pi d\theta_{\bf k}\,\sin\theta_{\bf k}\,
(x^2 + \sin^2\theta_{\bf k}\,\varphi^2)\,
\frac{1}{1+\cosh\left(\sqrt{x^2+\sin^2\theta_{\bf k}\,\varphi^2}\right)} \,\, .
\ee
In this case, the dominant contribution comes from the vicinities 
of the north and south pole of the Fermi sphere, where the gap vanishes. 
The gap behaves linear around these points, 
$\sqrt{\lambda_{{\bf k},1}} \simeq \theta_{\bf k}$. 
To obtain the leading order result in $\varphi$, we integrate
over the region where the quasiparticle energy is less than or 
of the order of the scale set by the temperature, i.e., 
$\sqrt{\lambda_{{\bf k},1}}\,\phi_0 \lesssim \pi\, T$, or, equivalently, 
$\theta_{\bf k} \lesssim \pi/\varphi$ (north pole of the Fermi sphere, 
the south pole yields the same result). Consequently,
\be
K(\varphi) \simeq \frac{3}{\pi^2}\int_0^\infty dx\,
\frac{1}{1+\cosh(x)}\int_0^{\pi/\varphi} d\theta_{\bf k}\,\theta_{\bf k}
(x^2 + \theta_{\bf k}^2\varphi^2) 
= \frac{5\pi^2}{4}\,\frac{1}{\varphi^2} \simeq \frac{12.34}{\varphi^2}\,\, .
\ee

\subsection{{\it A} phase}

In the {\it A} phase, we use Eq.~(\ref{KA}), i.e., in principle, we have to consider two different
angular gap structures. However, the small temperature behavior is dominated by the nodes of the
gap. Therefore, it is sufficient to consider only the second branch, 
$\lambda_{{\bf k},2}=(1-|\cos\theta_{\bf k}|)^2$,
\be
K(\varphi) =  \frac{3}{4\pi^2}\int_0^\infty dx\,\int_0^\pi d\theta_{\bf k}\,\sin\theta_{\bf k}\,
\left[x^2 + (1-|\cos\theta_{\bf k}|)^2\varphi^2\,\right]\,\frac{1}{1+\cosh\left(\sqrt{x^2+(1-|\cos\theta_{\bf k}|)^2\,\varphi^2}\right)} \,\, .
\ee
Contrary to the polar phase, the gap function behaves quadratically in the angular directions in the
vicinity of the nodes, $\sqrt{\lambda_{{\bf k},2}} \simeq \theta_{\bf k}^2/2$. Therefore, in order to be 
consistent with the estimate in the polar phase, we restrict the integral to the region 
$\theta_{\bf k} \lesssim \sqrt{2\pi/\varphi}$,   
\be
K(\varphi) \simeq \frac{3}{2\pi^2}\int_0^\infty dx\,\frac{1}{1+\cosh(x)}
\int_0^{\sqrt{2\pi/\varphi}} d\theta_{\bf k}\,
\theta_{\bf k}\,\left(x^2 + \frac{\theta_{\bf k}^4}{4}\,\varphi^2\right) 
= \frac{\pi}{\varphi} \simeq \frac{3.14}{\varphi}\,\, .
\ee


\begin{thebibliography}{99}

\bibitem{perry}
J.~C.~Collins and M.~J.~Perry,
Phys.\ Rev.\ Lett.\  {\bf 34}, 1353 (1975).

\bibitem{reviews}
K.~Rajagopal and F.~Wilczek,
hep-ph/0011333;
M.~Alford,
Ann.\ Rev.\ Nucl.\ Part.\ Sci.\  {\bf 51}, 131 (2001);
S.~Reddy,
Acta Phys.\ Polon.\ B {\bf 33}, 4101 (2002);
T.~Sch{\"a}fer,
hep-ph/0304281;
D.~H.~Rischke,
Prog.\ Part.\ Nucl.\ Phys.\  {\bf 52}, 197 (2004);
M.~Buballa,
Phys.\ Rept.\  {\bf 407}, 205 (2005);
H.-C.~Ren,
hep-ph/0404074;
M.~Huang,
Int.\ J.\ Mod.\ Phys.\ E {\bf 14}, 675 (2005);
I.~A.~Shovkovy,
Found.\ Phys.\  {\bf 35}, 1309 (2005).

\bibitem{bailin} 
D.~Bailin and A.~Love,
Phys.\ Rept.\ {\bf 107}, 325 (1984).

\bibitem{alford} 
M.~G.~Alford, K.~Rajagopal and F.~Wilczek,
Nucl.\ Phys.\ B {\bf 537}, 443 (1999).

\bibitem{weak-cfl1} I.~A.~Shovkovy and L.~C.~R.~Wijewardhana,
Phys.\ Lett.\ B {\bf 470}, 189 (1999).

\bibitem{weak-cfl2} T.~Sch\"afer,
Nucl.\ Phys.\ B {\bf 575}, 269 (2000).

\bibitem{neutrality}
K.~Iida and G.~Baym,
Phys.\ Rev.\ D {\bf 63}, 074018 (2001)
[Erratum-ibid.\ D {\bf 66}, 059903 (2002)];
M.~Alford and K.~Rajagopal,
JHEP {\bf 0206}, 031 (2002);
A.~W.~Steiner, S.~Reddy and M.~Prakash,
Phys.\ Rev.\ D {\bf 66}, 094007 (2002);
M.~Huang, P.~F.~Zhuang and W.~Q.~Chao,
Phys.\ Rev.\ D {\bf 67}, 065015 (2003);
F.~Neumann, M.~Buballa and M.~Oertel,
Nucl.\ Phys.\ A {\bf 714}, 481 (2003);
S.~B.~R\"uster and D.~H.~Rischke,
Phys.\ Rev.\ D {\bf 69}, 045011 (2004).

\bibitem{bcs} 
J.~Bardeen, L.~N.~Cooper and J.~R.~Schrieffer,
Phys.\ Rev.\  {\bf 108}, 1175 (1957).

\bibitem{shovkovy}
I.~A.~Shovkovy and M.~Huang, 
Phys.\ Lett.\ B {\bf 564}, 205 (2003);
M.~Huang and I.~A.~Shovkovy,
Nucl.\ Phys.\ A {\bf 729}, 835 (2003).

\bibitem{gCFL}
M.~Alford, C.~Kouvaris and K.~Rajagopal,
Phys.\ Rev.\ Lett.\  {\bf 92}, 222001 (2004).

\bibitem{cfl+mesons}
P.~F.~Bedaque and T.~Sch\"afer,
Nucl.\ Phys.\ A {\bf 697}, 802 (2002);
D.~B.~Kaplan and S.~Reddy,
Phys.\ Rev.\ D {\bf 65}, 054042 (2002);
A.~Kryjevski, D.~B.~Kaplan and T.~Sch\"afer,
Phys.\ Rev.\ D {\bf 71}, 034004 (2005);
A.~Kryjevski,
hep-ph/0508180;
T.~Sch\"afer,
hep-ph/0508190.

\bibitem{alford2}
M.~G.~Alford, J.~A.~Bowers and K.~Rajagopal,
Phys.\ Rev.\ D {\bf 63}, 074016 (2001).

\bibitem{spin1}
M.~Iwasaki and T.~Iwado,
Phys.\ Lett.\ B {\bf 350}, 163 (1995);
R.~D.~Pisarski and D.~H.~Rischke,
Phys.\ Rev.\ D {\bf 61}, 074017 (2000);
M.~G.~Alford, J.~A.~Bowers, J.~M.~Cheyne and G.~A.~Cowan,
Phys.\ Rev.\ D {\bf 67}, 054018 (2003).

\bibitem{schaefer}
T.~Sch\"afer, 
Phys.\ Rev.\ D {\bf 62}, 094007 (2000).

\bibitem{Schmitt:2002sc}
A.~Schmitt, Q.~Wang and D.~H.~Rischke,
Phys.\ Rev.\ D {\bf 66}, 114010 (2002).

\bibitem{Schmitt:2003xq}
A.~Schmitt, Q.~Wang and D.~H.~Rischke,
Phys.\ Rev.\ Lett.\  {\bf 91}, 242301 (2003).

\bibitem{andreas} A.~Schmitt,
Phys.\ Rev.\ D {\bf 71}, 054016 (2005).

\bibitem{vollhardt} 
D.~Vollhardt and P.~W\"olfle, 
{\it The Superfluid Phases of Helium 3} (Taylor \& Francis, London, 1990).  

\bibitem{Buballa:2002wy}
M.~Buballa, J.~Ho\v{s}ek, and M.~Oertel,
Phys.\ Rev.\ Lett.\  {\bf 90}, 182002 (2003).

\bibitem{KB} L.~P.~Kadanoff and G.~Baym,
{\it Quantum Statistical Mechanics}
(Benjamin, New York, 1962).

\bibitem{VoskSen} 
D.~N.~Voskresensky and A.~V.~Senatorov,
Sov.\ J.\ Nucl.\ Phys.\  {\bf 45}, 411 (1987)
[Yad.\ Fiz.\  {\bf 45}, 657 (1987)].

\bibitem{friman}
M.~Sch\"onhofen, M.~Cubero, B.~L.~Friman, W.~N\"orenberg and G.~Wolf,
Nucl.\ Phys.\ A {\bf 572}, 112 (1994).

\bibitem{kadanoff}
A.~Sedrakian and A.~Dieperink,
Phys.\ Lett.\ B {\bf 463}, 145 (1999);
Q.~Wang, K.~Redlich, H.~St\"ocker and W.~Greiner,
Phys.\ Rev.\ Lett.\  {\bf 88}, 132303 (2002);
Q.~Wang, K.~Redlich, H.~St\"ocker and W.~Greiner,
Nucl.\ Phys.\ A {\bf 714}, 293 (2003).

\bibitem{Brown:2000eh}
W.~E.~Brown, J.~T.~Liu and H.-C.~Ren,
Phys.\ Rev.\ D {\bf 62}, 054013 (2000).

\bibitem{Wang:2001aq}
Q.~Wang and D.~H.~Rischke,
Phys.\ Rev.\ D {\bf 65}, 054005 (2002).


\bibitem{jaikumar}
P.~Jaikumar, C.~D.~Roberts and A.~Sedrakian,
nucl-th/0509093.

\bibitem{iwamoto}
N.~Iwamoto,
Phys.\ Rev.\ Lett.\  {\bf 44}, 1637 (1980).

\bibitem{voskresensky}
D.N.\ Voskresensky, V.A.\ Khodel, M.V.\ Zverev, J.W.\ Clark,
Astrophys.J. {\bf 533}, 127 (2000).

\bibitem{schwenzer}
T.~Sch\"afer and K.~Schwenzer,
Phys.\ Rev.\ D {\bf 70}, 114037 (2004).

\bibitem{Yak} D.~G.~Yakovlev, A.~D.~Kaminker, O.~Y.~Gnedin and P.~Haensel,
Phys.\ Rept.\  {\bf 354}, 1 (2001).

\bibitem{voskresensky2}
D.N.\ Voskresensky,
Lect.\ Notes Phys.\  {\bf 578}, 467 (2001).

\bibitem{ren} 
W.~E.~Brown, J.~T.~Liu and H.-C.~Ren,
Phys.\ Rev.\ D {\bf 62}, 054016 (2000).

\bibitem{prl}
A.~Schmitt, I.~A.~Shovkovy and Q.~Wang,
Phys.\ Rev.\ Lett.\  {\bf 94}, 211101 (2005) [Erratum-ibid. {\bf 95}, 159902 (2005)].

\bibitem{Blaschke}
D.~N.~Aguilera, D.~Blaschke, M.~Buballa and V.~L.~Yudichev,
Phys.\ Rev.\ D {\bf 72}, 034008 (2005).

\bibitem{helicity-cond-mat}
C.\ Wu and S.-C.\ Zhang, 
Phys.\ Rev.\ Lett.\ {\bf 93}, 036403 (2004);
C.~M.~Varma and L.~Zhu,
cond-mat/0502344.


\end{thebibliography}
\end{document}